\documentclass[aps,twocolumn,showpacs,superscriptaddress,amsmath,amssymb]{revtex4}
.tfm scaled 1000 %  Kaligrafik
\newcommand{\bea}{\begin{eqnarray}}
\newcommand{\beq}{\begin{equation}}
\newcommand{\eea}{\end{eqnarray}}
\newcommand{\eeq}{\end{equation}}
\usepackage{graphicx}
\usepackage{graphics}
\usepackage{epsfig}
\usepackage{verbatim}
\usepackage{amsmath,amscd,amssymb,euscript,eufrak}

\def\vp{\varphi}          
\def\bbf#1{\boldsymbol{#1}}

\def\SUM#1#2{\mbox{$\sum\limits_{#1}^{#2} $}}

\def\begg#1{ \begin{gather} #1 \end{gather} }
\def\ra{\rangle}          \def\la{\langle}
\begin{document}
\title{
Self-organization of
charged particles in
circular geometry}
\author{R.G. Nazmitdinov}
\affiliation{
Departament de F{\'\i}sica,
 Universitat de les Illes Balears, E-07122 Palma de Mallorca, Spain}
\affiliation{Bogoliubov Laboratory of Theoretical Physics,
Joint Institute for Nuclear Research, 141980 Dubna, Russia}
\author{A. Puente}
\affiliation{
Departament de F{\'\i}sica,
 Universitat de les Illes Balears, E-07122 Palma de Mallorca, Spain}
\author{M. Cerkaski}
\affiliation{Department of Theory of Structure of Matter,
Institute of Nuclear Physics PAN, 31-342 Cracow, Poland}
\author{M. Pons}
\affiliation{
 Departament de F{\'\i}sica,
 Universitat de les Illes Balears, E-07122 Palma de Mallorca, Spain}
\begin{abstract}
The basic principles of self-organization of  one-component
charged particles, confined in  disk and  circular parabolic potentials,
are proposed.
A system of equations is derived, that allows us to determine equilibrium configurations
for an arbitrary, but finite, number of charged particles that are distributed over several rings.
Our approach reduces significantly the computational effort
in minimizing the energy of equilibrium configurations and
demonstrates a remarkable agreement with the values provided by molecular dynamics
calculations. With the increase of particle number $n>180$ we
find a steady formation of a centered  hexagonal lattice that smoothly transforms to
valence circular rings in the ground state configurations
for both potentials.
\end{abstract}
%\pacs{36.40.Wa,82.70.Dd,02.20.Rt,64.75.Yz}
\date{\today}
\maketitle
%05.65.+b 	Self-organized systems (see also 45.70.-n in classical mechanics of discrete systems)
%45.50.Jf 	Few- and many-body systems
%41.20.Cv 	Electrostatics; Poisson and Laplace equations, boundary-value problems
%64.70.kp 	Ionic crystals
%82.70.Dd 	Colloids
%64.75.Yz 	Self-assembly
%36.40.Wa 	Charged clusters
\section{Introduction}
There is an enormous interest in mesoscopic systems consisting of a finite number of
interacting particles in a confined geometry. It is well understood that
various phenomena, that are suppressed in a continuous limit, are brought about
by finiteness  and boundaries of these systems  \cite{bir}. Progress in modern technology
allows us to study such phenomena on the same scale, from Bose condensate with some 
thousand atoms to quantum dots with a few electrons, providing rich information about specific features of correlation effects in mesoscopic systems (see, for example, Ref.\cite{fin}).
Nowadays, many ideas and concepts introduced, in particular, in condensed matter physics
can be realized and analyzed with high accuracy as a function of  particle number
and boundary properties.

Long ago Wigner predicted  that electrons
interacting by means of Coulomb forces
could create a crystallized structure   in a three-dimensional (3D) space
at low enough densities and temperatures \cite{Wig}.
At these conditions the potential energy dominates over the kinetic energy and defines
equilibrium configurations of electronic systems.
This prediction initiated  various research lines in diverse branches of
physics and chemistry. In particular, the so called Coulomb clusters
that result from harmonic confinement of charged particles in two and three dimensions
attracted intensive attention, since they are relevant for the description of
cold ions in various traps, dusty plasmas, and many other systems.

At moderate number of particles ($\sim 10^3$) the properties of
spherical Coulomb systems may be analysed in terms of simple shell
models, in which the constituting particles create concentric spherical surfaces
called shells (see Ref.\,\onlinecite{cio1} and references therein).
The crystallization of a one-component plasma for a system size
up to $10^5$ ions, confined by a spherically
symmetric parabolic potential, induced by their mutual
Coulomb interaction, has been studied by means of molecular
dynamics (MD) simulations \cite{3MD}. It was found that the formation
of the {\sl bcc} lattice provides better ground state energies than
shell configurations for a number of ions $n\geq 2\times 10^4$.
Signatures of Wigner crystallization were also observed
in two-dimensional (2D) distributions
of electrons on the surface of liquid helium \cite{hel1}.
A phase transition, induced by magnetic field, from an electron
liquid to a crystalline structure has also been reported for a
2D electron plasma at a GaAs/AlGaAs heterojunction \cite{tr}.

In finite mesoscopic systems, with small number of particles, it is, however,
difficult to expect a phase transition. In these systems one observes 
crossovers rather than phase transitions. Therefore, the question of how the Wigner
crystallization may settle down in these systems is still an intriguing fundamental
problem. Leaving aside proper quantum mechanical descriptions, which due to symmetry
do not allow for particle localization (see, e.g., a discussion in Ref.\cite{lor})
even a classical picture needs further clarifications.
One needs to understand how a symmetry of
a restricted geometry affects physical and chemical
properties as a function of the number of interacting
charged particles.
Evidently, the decrease of system size
 places primary emphasis upon system boundaries.
 It appears that, in contrast to the 3D case,
a 2D system turns out to be more complicated for studies of  shell structure
and the onset of crystallization in systems with charged particles
of one species (see, e.g., a discussion in Ref.\cite{jer1}).
It is appropriate at this point to recall that, 
according to the Earnshaw's theorem \cite{et},
classical charges, confined in 2D hard wall, 
with logarithmic interparticle interaction  would end up
at the border of the potential (see also a discussion in Ref.\cite{lev}).

Meanwhile, the question of how charged particles arrange themselves
in a restricted planar geometry attracted a continuous attention for many
decades (for a review see \cite{1}). J.J. Thomson was the first to
suggest an instructive solution for interacting electrons,
reducing the 3D harmonic oscillator confinement
to a circular (2D) harmonic oscilator \cite{tom}.
He developed an analytical approach which enables to trace
a self-organization for a small number of electrons ($n<50$) in a family of rings (shells)
with a certain number of electrons in each shell (see details in \cite{tom1}).
Although the number of particles in outer and inner rings changes
as a function of the total number of electrons,
each shell is characterised by a certain discrete symmetry.
In other words, $n$ point charges, located on the ring,
create equidistant nodes on this ring with the angular step $\alpha=2\pi/n$.
Similar shell patterns have been found much later by means of Monte--Carlo (MC)
calculations \cite{loz,bol} for charged particles (ions and electrons)
confined by a 2D parabolic and hard-wall potentials
(see, e.g., Refs.\,\cite{pet1,pet2} for a systematic analysis
of small number of charges $n\leq52$). Ground states of a few
electrons in various polygons have also been analysed by means of unrestricted
Hartree--Fock and density-functional theory calculations (e.g., \cite{ras,jap}).
Structures of polygonal patterns, similar to those obtained with an effective harmonic oscillator
confinement, have been observed in experimental
measurements \cite{epl,prb79}.
In many cases, the polygonal pattern of equally charged particles
is sufficiently regular.

From the above analysis, based on MC and MD calculations \cite{wor} for a relatively small
number of charged particles, it follows that the number of
stable configurations grows very rapidly with the number of particles.
There are many local minima that have energies very close to the global minimum.
These metastable states with lower (or higher) symmetry are found with much higher
probabilities than the true ground state \cite{am,toni}. This picture is akin to a liquid-solid transition,
when a rapid cooling gives rise to a glasslike disordered solid rather than
a crystal with lower energy.
In this case various simulations techniques are too labour-intensive to be chosen
for a thorough analysis of the system with increasing particle number.
Evidently, a search procedure for the ground state of
such systems becomes of paramount importance.
One of the major aims of our paper is to provide an effective
 semi-analytical approach that enables us to describe
ground state properties of charged particles in
a circular  potential as a function of particle number with a good accuracy.
In order  to avoid a large admixture of metastable states with the ground state
we consider particles interacting by means of the Coulomb forces at zero temperature.
Although we consider  classical systems,
our approach could shed
light on the nature of self-organization of colloidal particles in organic solvents,
charged nanoparticles absorbed at oil-water interfaces, electrons trapped on
the surface of liquid helium or ionized plasmas. It is pertinent to note, however, that
for such systems the interaction could be more complex
than the one considered in our paper (cf. \cite{lev1}).

For completeness we mention that similar problems have been studied
in a continuous limit \cite{zar,koul,mug}. In Ref.\,\onlinecite{zar} a classical hydrodynamic approach
has been developed to analyze magnetoplasmonic excitations in electron quantum dots.
This approach can be viewed as the simplest density-functional theory of a confined
electron gas with electronic interactions treated in the Hartree approximation.
A general trend of the density distribution in disk and parabolic potentials was considered
in the framework of elasticity theory \cite{koul,mug}.
These approaches are, however, unable to provide a detailed description of the
shell structure for finite number of charged particles $(n \leq 1000)$.
An asymptotic description of Coulomb systems
confined by radially symmetric potentials in two and three dimensions is
discussed in Ref.\,\onlinecite{c2}. Although this approach is akin
to ours, it lacks the detailed analysis provided in the present paper.

In order to check the validity of our theoretical approach,
we also develop and perform MD calculations and compare
our predictions with the MD results.
In Ref.\,\onlinecite{sup} the reader can find our results corresponding to MD
and the semi-analytical approach for $n\leq 200$ particles with a parabolic confinement.

The structure of the paper is as follows.
In Sec.\,II we recapitulate the basic ideas of our approach,
briefly discussed in Ref.\,\onlinecite{cnp} for disk geometry,
and obtain the analytical formula for  the ring-ring interaction.
Sec.\,III is devoted to the extension of our approach for the circular parabolic potential and
a comparison of the results obtained under disk and parabolic confinements.
In Sec.\,IV we discuss the basic ideas of our MD approach  and compare
the results with those obtained within our semi-analytical approach.
The main results of our analysis are summarized in Sec.\,V.
 In two Appendixes we provide technical details and prove some statements
that have been taken for granted in Ref.\,\onlinecite{cnp}.

\section{Coulomb interaction and cyclic symmetry}
\subsection{Model system}

We study a system of $n$ identical charged particles with
 Coulomb interactions in a 2D confining potential. The Hamiltonian reads
 \begin{equation}
 \label{ham}
 H=\sum_{i=1}^nV(r_i)+\alpha\sum_{i<j}^n\frac{1}{|\vec{r}_i-\vec{r}_j|}\,,
 \end{equation}
 where $r_i=|r_i|$ is the particle distance to the center of the confining
 potential, and $\alpha=e^2/4 \pi \varepsilon_0 \varepsilon_r$ characterizes the interaction strength in the host material. Although we choose electrons as an example, the charged 
 particles could be ions as well.

We consider two confining potentials:

1) a hard-wall (disk) confinement
\begin{equation}
\label{17}
V(r) \,=\,
\left \{
\begin{array}{l}
0\,,
\qquad
\,\,\,\,\,\,  r<R \\
\infty\,,\qquad
\,\,\,\,\,\, r\geq R \\
\end{array},
\right.
\end{equation}

2) a circular parabolic potential
\begin{equation}
\label{hop}
V(r)=\frac{1}{2}m\omega_0^2r^2\,.
\end{equation}

As discussed above, in many numerical simulations,  the interaction
between a finite number of charged particles leads to the formation of shells in parabolic
and disk potentials. These shells consist in a family of rings at different radii, $r_i$,
which are occupied by a specific number of particles. In each shell $n_i$  point charges
create equidistant nodes on the ring, with an angular spacing $\alpha_i=2\pi/n_i$.
Although a similar pattern is obtained for the parabolic and disk potentials,
the distribution of particles over rings is very different in two cases.
Below we  will attempt to shed light on the similarity and difference in
the self-organization of charged particles in both systems with aid of the semi-analytical
approach. The key ingredient of that approach is an effective method
for evaluating the various ring-ring energies.
The method can be applied to any interaction characterized by the cyclic symmetry.

\subsection{Interaction between  two rings}

We recall that the Coulomb energy of  $n$ unit charges $e$, equally distributed
over a ring of radius $r$, has the following form \cite{tom}:
\bea
\label{Ering}
&& E_n(r)=\frac{\alpha}{2\,r}\sum_{i=1}^{n-1}\sum_{j=i+1}^{n}\frac{1}{\sin\frac{\pi}{n}(|i-j|)}=
\frac{\alpha\, nS_n}{4\,r}\,,\\
\label{sn}
&& S_n=\sum_{k=1}^{n-1}\frac{1}{\sin \frac{\pi}{n}k}\,.
\eea
Below, for the sake of discussion, we use $\alpha=1$, unless stated otherwise.
For increasing number of particles several rings build up 
(e.g., Refs.\cite{loz,bol,pet1,pet2,wor,pet3}).
To compute the total energy we need to add the contribution that is due to
ring-ring interactions, which is absent in the Thomson model.
This is the first basic ingredient of our approach.

The interaction between two rings with $n$ and $m$ point charges can be expressed as
\bea
\label{en}
 E_{nm}(r_1,r_2,\psi)& =&  \sum_{i=1}^n\,\sum_{j=1}^m\,\epsilon (r_1,r_2,\psi^{nm}_{ij}+\psi)\nonumber\\
  &=& {\rm G}\times \sum_{k=1}^{\rm L}\, \epsilon (r_1,r_2,\psi_k+\psi)\,,\\
 \label{eijnm}
 \epsilon (r_1,r_2,\theta) & = &(r_1^{2}+r_2{^2}-2\,r_1\,r_2\,\cos \theta)^{-1/2}\,,
\eea
where $ \psi^{nm}_{ij}= 2 \,\pi(i/n -j/m)$ and $\psi$ stands for the relative angular
offset between both rings. Here,
$\{\psi_k= \Delta\times k, k=1,\ldots, L\}$ and $L\equiv {\rm LCM}(n,m)$,
$G\equiv{\rm GCD}(n,m)=n\times m/L$ are the least common multiple and
greatest common divisor of the numbers $(n,m)$, respectively. The ring-ring energy is
a periodic function with a $\Delta=2\pi / L$ periodicity. In turn,
this result shows that these kind of functions are invariant under angle transformations
corresponding to the cyclic group of $L$ elements.
The proof of this result is given in Appendix \ref{red}.

In virtue of the fact that
the ring-ring interaction is an even periodic function in the angle $\psi$, it can be expressed
by means of a cosine Fourier series
\beq
\label{fser}
E_{nm}(r_1,r_2,\psi)=\la E_{nm} \ra \! +\! \SUM{\ell=1}{\infty}C_{\ell nm}(r_1,r_2)\cos(\ell L \psi).
\eeq
The average value is obtained by integrating in $\psi$, and, using Eq.\,(\ref{en}), we have
\bea
\label{aven}
\la  E_{nm} \ra=\frac{1}{2\pi}\int_0^{2\pi}\!\!\!\!\! d\psi \, E_{nm}(r_1,r_2,\psi) \nonumber\\
=\frac{G}{2\pi}\SUM{k=1}{L}\int_{0}^{2\pi} \!\!\!\!\! d\psi \,
\epsilon(r_1,r_2,\psi_k+\psi)\, .
\eea

All terms in the sum Eq.\,(\ref{aven}) give the same contribution, and we obtain
in terms of the complete elliptic integral of first kind  (see Ref.\,\onlinecite{AS}, p.\,590)
\beq
\la  E_{nm} \ra=\frac{2nm}{\pi r_>(1+t)} K(4 t/(1+t)^2) =2nm\frac{K(t^2)}{\pi r_>}\, .
\label{aven2}
\eeq
Here, we introduced notations: $r_>=\max(r_1,r_2)$, $r_<=\min(r_1,r_2)$, $t=r_</r_>$;
and  used the symmetry property
$K(4t/(1+t)^2)  = (1 + t)\,  K(t^2)$.
It is noteworthy that the average value  $\la E_{nm} \ra$ is exactly the interaction energy
between homogeneously distributed $n$ and $m$
charges over the rings.

In a similar way, the Fourier coefficients corresponding to the fluctuating part of the energy,
$\Delta\,E_{nm} = \SUM{\ell=1}{\infty}C_{\ell nm}(r_1,r_2)\cos(\ell L \psi)$,
are given by
\begg{
C_{\ell nm}(r_1,r_2)=\frac{1}{\pi}\int_0^{2\pi} \!\!\!\!\! d\psi \, \cos(\ell L \psi) E_{nm}(r_1,r_2,\psi)\nonumber\\
=\frac{nm}{\pi}
\int_0^{2\pi} \!\!\!\!\! d\psi \, \frac{\cos(\ell L\psi)}{\left[r_1^2+r_2^2-2r_1r_2\cos \psi \right]^{1/2}}\,.
\label{clm}
}
We have derived the following analytical expression for this integral
$(\ell\geq 1$; see details in Appendix \ref{coef})
\bea
\label{ff1}
C_{\ell nm}(r_1,r_2) & = &
\frac{2 n m}{r_>}\, \frac{(2\ell L-1)!!}{(\ell L)!}\left( \frac{t}{2} \right)^{\ell L}
   \times\\
    &&\times {}_2 F_{\mspace{-1mu}1} \Bigl( 1/2 \,,\ell L+1/2 \,;\ell L +1 \,;\, t^2\Bigr)\nonumber\,,
\eea
which is quite convenient for evaluation with symbolic algebra packages.

In particular, at $\ell=0$ one  obtains the result (\ref{aven2}):
\bea
&&  {}_2 F_{\mspace{-1mu}1}\:\Bigl( 1/2 \,,\, 1/2 \,;\, 1 \,;\, z^2\Bigr)  =
\frac{2}{\pi}\: K(z^2)  \nonumber\\
&& \Rightarrow C_{0nm}(r_1,r_2)=2n m\frac{K(t^2)}{\pi r_>}\,.\nonumber
\eea

\section{Ground state  configurations: semi-analytical approach}
Before we tackle the problem of self-organization of charged particles confined
in the parabolic potential, it is useful  to review briefly the results obtained
for the hard-wall (disk) potential.

\subsection{Hard-wall confinement}
In this case, the total energy is defined as
\begin{equation}
\label{Ec}
 E_{\rm tot}( {\bf n},{\bf r},{\bbf{\vp}} ) =
     \sum_{i=1}^{p} E_{n_i} +
     \sum_{i=2}^{p}\sum_{j=1}^{i-1}\, E_{n_i n_j}(r_i,r_j,\vp_{ij})\,.
\end{equation}

Here, ${\bf{n}}\!=\!(n_1,\ldots,n_p)$ is a partition of the total number $n$ on
$p$ rings with radii ${\bf{r}}\!=\!(r_1,\ldots,r_p)$ and
offset angles between different rings $(i<j=2,3,\ldots,p)$:
$\bbf{\vp}\!=\!(\vp_{12},\ldots,\vp_{1p},\vp_{23},\ldots )$.
We assume $R=r_1=1 > r_2 > \cdots > r_p$.
The numerical analysis \cite{cnp} demonstrates
that
 \beq
        E_{n_i,n_j}(r_i,r_j,\vp_{ij})  \simeq   \langle E_{n_i,n_j} (r_i,r_j)\rangle
\eeq
holds for $n\leq 2000$ with a high accuracy.  Therefore, we
neglect the dependence on
the relative angles $\vp_{ij}$ , {\em i.e.}, the fluctuating term $\Delta\,E_{n_i ,n_j}$.
%(\ref{res}).
The total energy of $n$ charged particles in a disk of radius $R$ is then
$E_{\rm tot}( {\bf n},{\bf r},\bbf{\vp} ) \simeq{\cal E}_{\rm avg}({\bf{n},\bf{r}})$
with
\beq
\label{Ecoll}
{\cal E}_{\rm avg} ({\bf{n},\bf{r}}) =  \SUM{i=1}{p}\,n_i\frac{S_{n_i}}{4r_i}+
\frac{2}{\pi} \SUM{i<j}{p} \,n_i\,n_j\,\frac{K ( (r_j/r_i)^2) }{r_i}\,.
\eeq

The equilibrium configuration of particles
can be obtained by minimizing Eq.\,(\ref{Ecoll}) with respect to $(p,\bf{n},\bf{r})$,
{\em i.e.}, finding the partition corresponding to the lowest total energy.
For a given partition, the set of equations that
determines the optimal radii $r_i$ is
\begin{equation}
\label{minr}
{\cal F}_i =\frac{\pi}{2} ( r_i/n_i)\frac{d}{dr_i}{\cal E}_{\rm avg}({\bf{n},\bf{r}})=0\,,
\quad  i=2,\ldots,p\,,
\end{equation}
where
\bea
\label{req1}
{\cal F}_i&=&r_i^{2} \sum_{j=i+1}^{p}{}\, \frac{n_j\,{\rm E}( (r_j/r_i)^2)}{r_j{^2} - r_i^2}-
\frac{\pi}{8}S_{n_i}+\\
 &+&  r_i\,\sum_{j=1}^{i-1}{} n_j\,\Bigl( \frac{ r_j\,{\rm E}((r_i/r_j)^2)}{r_j^2 - r_i^2}
 - \frac{ {\rm  K}((r_i/r_j)^2)}{r_j}\Bigr) \,. \nonumber
\eea
Here  ${\rm K}={\rm X_{-1}} \, ({\rm E}={\rm X}_1)$  are complete elliptic integrals of first (second) kind:
${\rm X}_p(x)  = \int_0^{\pi/2} dt\,(1-x\,\sin^2 t)^{p/2}$.
A few standard iterations  of Eqs.\,(\ref{minr}) suffice to reach an optimal energy value (\ref{Ecoll})
for a given partition. By sweeping a grid of different partitions, one can readily find the
lowest energy configuration for any fixed $n$.

\subsubsection{Structure of magic configurations}
Here,  we consider "magic configurations" for $n\leq 395$ charges as an example.
The minimization of energy with respect to the ring's partition numbers $\bf{n}$
leads to the following configurations:
\bea
\begin{array}{ccccccccccc}
 11: & 11   \\
 29:&  6 &     23 \\
 55: &5 & 13 &  37 \\
90:& 5 &  12 & 20 & 53 \\
135:& 5 & 12 & 19 & 29 & 70 \\
186:& 5 &  12 & 19 & 26 & 37 & 87 \\
 246:& 5 &  12 & 18 & 25 & 34 & 46 & 106 \\
 316:& 5 & 11 &  18 & 25 & 33 & 42 & 56 & 126 \\
394:&  5 &  11 &  18 &  25 & 32 & 40 & 50 & 66 & 147
 \end{array}
 \eea
These configurations are characterized by complete $p$ shells, {\em e.g.}, 
$p\!\!=\!\!9$ for $n\!\!=\!\!394$.
Our results provide an approximate formula for the number $p$ associated with a given particle
number
\beq
p_{H}\simeq[\sqrt{n}/2]\,.
\eeq
Here we introduce the subindex "H" associated with a hard-wall potential.

If one electron is added to the configuration with complete $p$ shells,
the {\it centered hexagonal configurations} (CHCs) start to appear, with a number
of electrons $6\,p,\ p=1,2,3\ldots$, surrounding one particle at the centre.
This tendency manifests clearly, starting from $n\geq56$, {\em i.e.}, we have
 \bea
\begin{array}{ccccccccccc}
12: & 1 &   11 \\
 30:&  1 &   6 &     23 \\
 56: &\bf{1} &  \bf{6}& \bf{12} &  37 \\
92:&   \bf{1}  &  \bf{6} &  \bf{12} & 20 & 53 \\
136:&  \bf{1}&  \bf{6} & \bf{12} & 19 & 28 & 70 \\
187:&  \bf{1} & \bf{6} &  \bf{12} & \bf{18} & 26 & 37 & 87 \\
 248:& \bf{1} &  \bf{6} &  \bf{12} & \bf{18} & 25 & 34 & 46 & 106 \\
 317:&  \bf{1} &  \bf{6} & \bf{12} &  \bf{18} & 25 & 32 & 42 & 55 & 126 \\
395:&  \bf{1} & \bf{6} &  \bf{12} &  \bf{18} &  \bf{24} & 32 & 40 & 50 & 65 & 147,
\end{array}
\eea
with the formation of new shells and a sequence of particles in the CHC which
is a characteristic property of the centered hexagonal lattice (CHL). Note that for
$n\!=\!92 \, (1,6,12,20,53)$,
$n\!=\!248 \, (1,6,12,18,25,34,46,106)$, the onset of the $6\,p$ rule needs one more particle.
After formation with each new shell, these recurrent internal CHCs persist till the addition of more
particles results in a sequentially increasing occupation of the inner ring, $n_1\!\!=\!\!2,3,4,(5)$,
and back again. As expected, the span of $n$ values which exhibit this internal CHC increases with $p$.

\begin{figure}[bth]
\vspace{-1.2cm}
\hspace{-0.91cm}
\includegraphics[scale=.232]{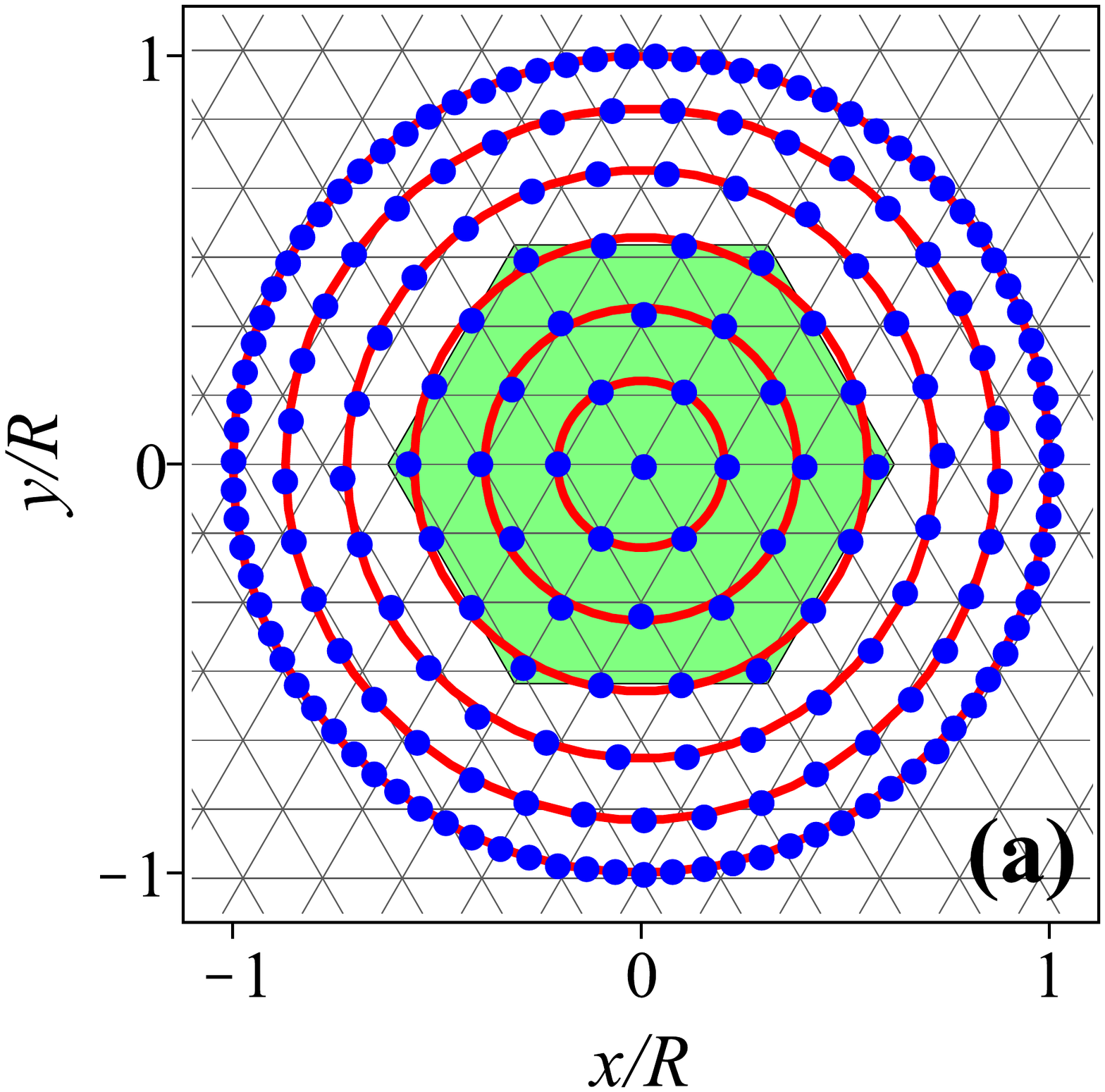}
\hspace{-.45cm}
\includegraphics[scale=.23]{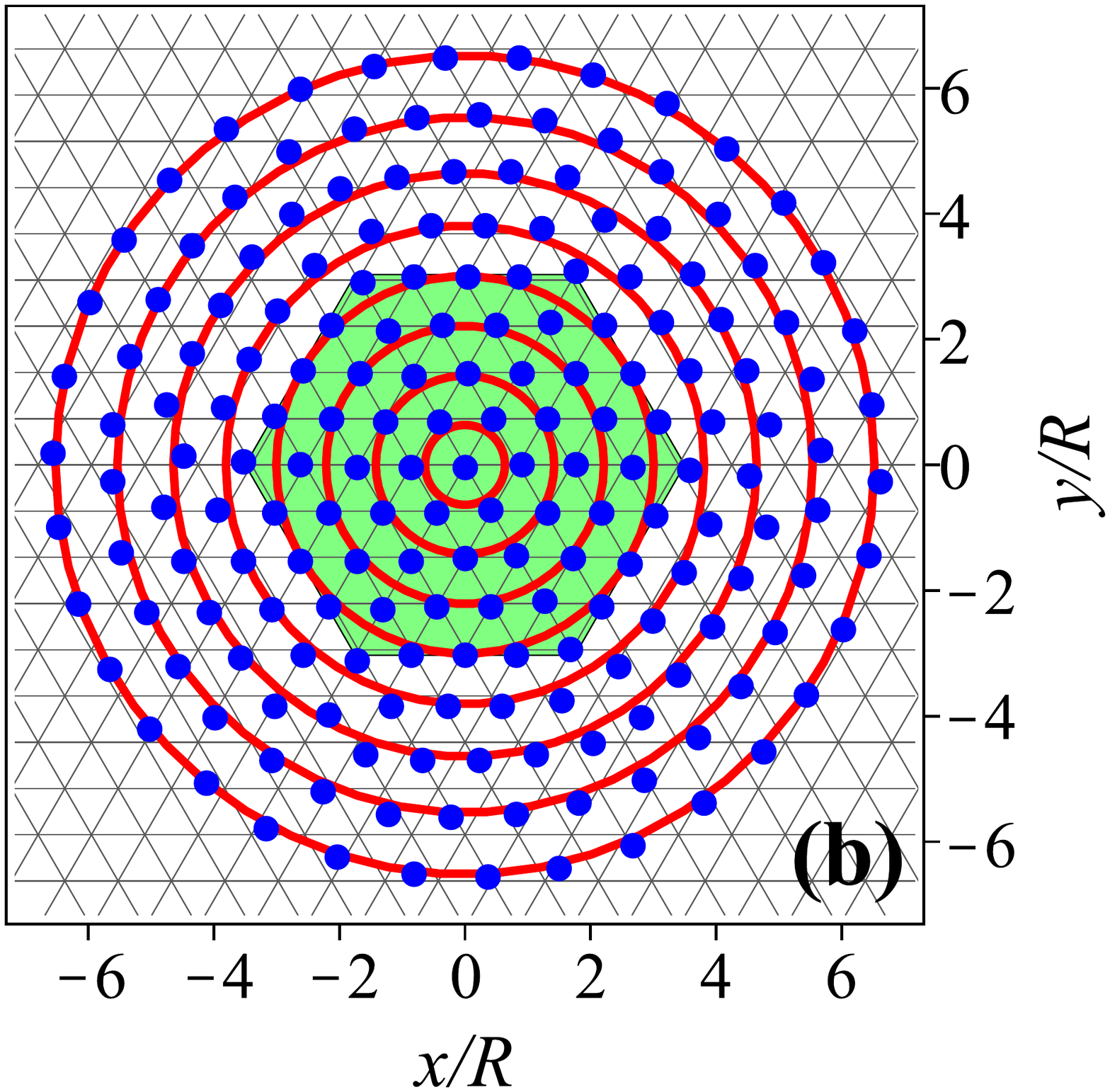}
\vspace{-1.2cm}
 \caption{(Color online)
  Structure of equilibrium configurations for the disk geometry (a)
  and for the circular parabolic potential (b) for $n=187$. In both systems there are
  internal ({\it core}) rings corresponding to the CHL (green region).
  Each shell in the {\it core} contains a family of circles
  with radii $R_{k \ell}$ and particle numbers $n_{\rm ring}=6\, p$ (see text).
 The numerical solution of Eqs.\,(\ref{minr}), (\ref{req}) (rings) are compared with
 the MD results (dots). The calculated energies
${\cal E}_{\rm avg}= 23652.9/6099.5$ for disk/parabolic potentials
are in a remarkable agreement with the corresponding MD results
${\cal E}_{\rm MD}= 23652.2/6098.8$.
 The {\it core} region with
 ${\{\bf 1}, {\bf 6}, {\bf 12}, {\bf 18}, ({\bf 24})\}$ particles, exhibits a hexagonal pattern.
 The {\it valence} shells contain $87,37,26$ and
  $34, 34, 34, 24$ particles with an almost perfect circular structure
  for the disk geometry and the circular parabolic potential, respectively.}
 \label{ss}
 \end{figure}

This fact can be understood by considering the arrangement of the
CHC points,  $\vec{x}_{k,\ell}=k\vec{a}_1+\ell \vec{a}_2$, given by integers $k, \ell$ and
the two primitive Bravais lattice vectors $\vec{a}_1=a\,(1,0)$ and
$\vec{a}_2=a\,(1/2,\sqrt{3}/2)$, where $a$ is the lattice constant. The $n_p=6 \, p$
sites in the $p\!-\!{\rm th}$ shell are organised
in different circular rings with radii $R_{k \ell}=a\,\sqrt{k^2+ \ell^2+k\,\ell}$,
where $p=k + \ell$ and $0\leq \ell \leq k$,
containing either 6 (if $\ell=0, k$) or 12 (otherwise) particles (see Fig.\ref{ss}a).
Up to $p=7$ all these radii are well ordered within and between
successive shells, and the model we presented groups them in a single
circular shell $n_{\rm ring}=6\, p$.
Beyond the seventh shell, however, rings start to overlap ({\em e.g.}, $R_{7,0} > R_{4,4}$),
ultimately distorting this sequence as they depart from the centre. In other words, the CHC becomes
broken giving up the reflection symmetry. The comparison of the equilibrium energies
and configurations, calculated by means of our semi-analytical model and the MD,
can be found in Ref.\,\onlinecite{sup}. We return to this point in Sec.IVB.

The systematic manifestation of the CHL with the increase of particle number
$n\geq 187$  can be interpreted as the onset of the centered
hexagonal crystallization in the disk geometry.
We recall that for {\it infinite} systems the hexagonal lattice has the lowest energy of
all two-dimensional Wigner Bravais crystals \cite{mar}.
However, in our finite system a crossover takes place from a
centered hexagonal {\it lattice} to {\it ring} localization at large $n$ with the
approaching to disk boundary.

Thus, we have found a cyclic self-organization for finite number of charged particles
confined in a disk geometry  (cf. Refs.\,\onlinecite{pet2,mug}).
For centered configurations particles
localize in shells, where each $p\!-\!{\rm th}$ internal shell consists in particles distributed over
a regular hexagon which is delimited by inscribed and circumscribed rings.
This CHC pattern is replaced by a ring organization when approaching the boundary.
A natural question arises: how  general is this type of organization?
Does one find a similar self-organization for other types of circular confinement?
As a next example, we consider charged particles confined by an external parabolic potential.

\subsection{Parabolic potential}

In the case of a circular parabolic potential (\ref{hop})
the Hamiltonian (\ref{ham}) obeys a scaling law. We can express the energy and the coordinates
in the following units
\beq
\label{un}
R=\left(\frac{\alpha}{\beta}\right)^{1/3}\,,\quad
e_0=\alpha^{2/3}\beta^{1/3}\,,\quad
\beta=m\omega_0^2\,,
\eeq
where $\alpha=e^2/4 \pi \varepsilon_0 \varepsilon_r$.
In such units the Hamitonian (\ref{ham}) can be written in the following form
\begin{equation}
\label{sch}
{\cal H}=\frac{H}{e_0}=\frac{1}{2}\sum_{i=1}^n x_i^2+
\sum_{i<j}^n \frac{1}{|\vec{x}_i-\vec{x}_j|}\,,
\end{equation}
where $\vec{x}_i=\vec{r}_i/R$. In the form (\ref{sch}) the Hamiltonian
does not depend on a particular value of
the confinement frequency $\omega_0$ or the interaction strength $\alpha>0$.
Therefore, the following analysis describes
universal properties for this confining potential.

\subsubsection{Structure of magic configurations}
To compare the results of our model with those already obtained in the disk geometry,
we also neglect the energy fluctuating term in Eq.\,(\ref{fser}).
Here, we consider the ground-state configurations for $n\leq200$ (see details
in Ref.\,\onlinecite{sup}).
In contrast to the disk geometry, we are forced to find the maximal radius
$r_1$ as a function of the particle number.
In the scaled  variables, within our approximation, we have for the total energy
\beq
\label{paren}
{\cal E}_{\rm avg}^S={\cal E}_{\rm avg} ({\bf{n},\bf{x}})+\frac{1}{2}\SUM{i=1}{p}\,n_i x_i{^2}\,.
\eeq
Here, the function ${\cal E}_{\rm avg} ({\bf{n},\bf{x}})$ is defined by Eq.\,(\ref{Ecoll}), where
$r_i$ is replaced by $x_i=r_i/R\,,x_i>x_{i+1}$.
We also introduce the index "S" associated with the parabolic (soft) potential.
The minimal energy configurations are obtained from the solution of the
system of equations
\beq
\label{req}
 {\cal F}_i^S={\cal F}_i+\frac{\pi}{2}x_i^2=0\,, \qquad i=1,\ldots,p\,,
 \eeq
 where ${\cal F}_i$ is determined by Eq.\,(\ref{req1}) with
 $r_i$ replaced by $x_i=r_i/R$.

The numerical results determine the  "magic configurations" with complete $p$ shells.
The structures of these configurations are different from those of the hard-wall potential:
\bea
\begin{array}{ccccccccccc}
 5: & 5   \\
 15:&  5 & 10 \\
 32:&  5 & 11 &  16 \\
 52:&  5 & 11 & 16 & 20 \\
 79:&  5 & 11 & 17 & 22 & 24 \\
111:& 5 & 11 & 17 & 22 & 27 & 29 \\
148:& 5 & 11 & 17 & 22 & 28 & 32 & 33 \\
190:& 5 & 11 & 17 & 23 & 28 & 33 & 36 & 37
\end{array}
 \eea

This difference manifests also in the number of rings for the same total number of particles, $n$.
Our results yield the following approximate formula for the number of shells
\beq
p_S\simeq [\sqrt{n/2}-1]\,.
\eeq
Compared to the hard-wall potential, there are
more rings in the parabolic confinement (see also Fig.\,\ref{ss}).
Nevertheless, similar to the disk potential, the addition of one electron
starts developing the CHC, with internal shell occupations $6\,p$, $p=1,2,3...$.
In contrast to the hard-wall case, the onset of this $6\,p$ rule is slightly
{\it delayed} after the beginning of each new shell. This is reflected in the
following results

\bea
\begin{array}{ccccccccccc}
  6: & 1 & 5   \\
 17: & {\bf 1} & {\bf 6} & 10 \\
 35: & {\bf 1} & {\bf 6} & {\bf 12} &  16 \\
 56: & {\bf 1} & {\bf 6} & {\bf 12} &  17 & 20 \\
 84: & {\bf 1} & {\bf 6} & {\bf 12} & {\bf 18}  & 22 & 25 \\
116: & {\bf 1} & {\bf 6} & {\bf 12} & {\bf 18} & 23 & 27 & 29 \\
155: & {\bf 1} & {\bf 6} & {\bf 12} & {\bf 18} & {\bf 24} & 28 & 32 & 34 \\
198: & {\bf 1} & {\bf 6} & {\bf 12} & {\bf 18} & {\bf 24} & 29 & 33 & 37 & 38 \\
 \end{array}
 \eea

Note that, due to the increase in energy caused by the parabolic potential,
less particles are located in this case on the rings close to the boundary.
In fact the corresponding scaling for the outer shell occupations are
\bea
\label{Scaln1}
n_1^H & = & \Big[ 2.795\, n^{2/3} - 3.184  \Big]\,,\\
n_1^S & = & \Big[ 0.2423\, n^{2/3} + 6.229\, n^{1/3} - 6.375  \Big]\nonumber\, .
\eea
These values are obtained from the fitting of the results
at the range $n=2-400(300)$ for the hard wall (parabolic) potential.
The analysis of this systematic data for the hard wall confinement also indicates that occupations
for subsequent shells are quite accurately predicted ($\pm 1$) by our model.
In particular, the  second shell occupation is fitted by
\beq
\label{n2}
n_2^H=\Big[ 1.351\, n^{2/3} - 6.566  \Big].
\eeq

In the disk geometry
the energy minimization distributes a large part of particles over the perfect
circular boundary. This group of charges stipulate the intrinsic ring organization
in this geometry. Considering the same number of particles and system size
($r^S_{\rm ext}=r^H_{\rm ext}$), in the parabolic confinement,
obviously $E^S_{\rm Coul} > E^H_{\rm Coul}$ since the equilibrium configuration
$({\bf{n}},\,{\bf{r}})_H$ is the one that minimizes Coulomb energy. Moreover,
as a consequence of the virial theorem ($E^S_{\rm Coul}=2\,E_{\rm HO}$), the total energy
$E^S=3\,E^S_{\rm Coul}/2$ is also bigger.
In order to distribute the larger amount of energy in this case
the system requires  additional shells, absent in the disk geometry,
to equilibrate the configuration. As a consequence, in the parabolic confinement
the distribution of particles over rings is less inhomogeneous
as compared to the disk geometry (see also Fig.\,\ref{ss}).
In turn, this  feature favours the formation of a more extended CHL.

In general, the increase of particle number in a new shell disintegrates slowly
 the CHL in both systems. As soon as a particle appears at the
 center, it gives rise to the CHL again.
 Below, for the sake of discussion we name our semi-analytical approach as
 the circular model (CM).

\section{Molecular dynamics}

\subsection{Basic approach}

To test our results,  we consider in the following both harmonic
$V_{\rm ext}(r)=1/2\,m\,\omega_0^2\, r^2$ and hard-wall $V_{\rm ext}(r)=\infty \Theta (r-R) $
confinements. Finding the absolute energy minimum, $E_{\rm gs}$ of the Hamiltonian (\ref{ham})
is a non trivial task. The density of stable states near $E_{\rm gs}$ grows exponentially
with the number of charges. Several global optimization techniques have been
extensively used to this aim. Metropolis simulated annealing, with temperature $T$ as a control
parameter, is particularly effective for short range forces. This method works on the basis
of acceptance/rejection of a proposed change in the particle positions (and corresponding
change in energy $\Delta E$)  with probability $p(\Delta E)=\exp(-\Delta E/k_{\rm B}\,T)$.
Random small changes over a single particle at a time are needed in order to guarantee
a reasonable acceptance  ratio, $r$. Usual practice adjusts the maximum change
at any given $T$ as to have $r\sim 1/2$.

In our pursuit of the exact ground state configurations we have used a
different approach, based on a quenched
molecular dynamics algorithm. The method evolves the particle positions by integrating the equations
of motion for the Hamiltonian (\ref{ham}) and adding a friction term which provides a controllable
and smooth quenching of velocities
\beq
\label{EqMD1}
m\,\vec{r}_i^{\ ''} = -\nabla_i\, V(\vec{r}_i) - b_{\rm f} \,\vec{r}_i^{\ '} \, .
\eeq
Here $V(\vec{r}_i)=V_{\rm ext} (r_i)+ \alpha\, \sum_{j\ne i} 1/r_{i\,j}$
includes the external potential plus the Coulomb terms, and $b_{\rm f}$ is the parameter
controlling the quenching of velocities. Below we discuss in detail our MD approach,
using the circular parabolic potential as an example. A few results of the MD calculations
for disk  are presented in Refs.\,\onlinecite{cnp,toni,sup}.

With the aid of the units (\ref{un}), writing
$\vec{r}=R\,\vec{x},\ E=e_0\,{\cal E}_{\rm MD}$ and $t=\tau/\omega_0$,
the dynamical equations become independent of the scaling parameters
for the harmonic potential
\beq
\label{EqMDad}
\ddot{\vec{x}}_i = -\vec{x}_i +
\sum_{j\ne i}^N \frac{\vec{x}_i - \vec{x}_j}{|\vec{x}_i - \vec{x}_j|^3}
 - \nu_{\rm f} \,\dot{\vec{x}}_i \, ,
\eeq
where $\dot{f}\equiv df/d\tau$ and $\nu_{\rm f}$ is the scaled friction parameter.
Given initial positions, the system (\ref{EqMDad}) is evolved by standard centered 3-point
derivative formulas, until forces over each particle are within a tolerance value
(typically $|\ddot{\vec{x}}_i| < 10^{-6}$).
An advantage of this method over the Metropolis algorithm is that it produces a sensible
global move at each iteration, thus requiring less simulation time.
The performance of this method for obtaining ground
state configurations is better than with usual Monte Carlo simulations, provided a sufficient number
of initial conditions are tried. In fact, one of the common errors when using any of these algorithms
is to be short in the number of trials and
getting as a result an excited state configuration. To avoid as much as possible this scenario,
we have included a systematic search of the $n$-particle configuration based on the lowest energy results for
$n\!-\!1$ particles. For each of these configurations, the new particle is placed at different random positions.
The new distinct configurations for the $n$ system are stored and ordered in energy to be used as
starting points for the $n\!+\!1$ system. This strategy is suitable for a systematic search of ground state
configurations in an ascending chain of $n$ charge particles.

\subsection{Search strategy for a single case}

In the case of a single (large) $n$ confined system,
the lowest lying energy results of the circular model, with much less degrees of freedom,
provide a convenient way to feed the MD analysis with sensible initial guesses.
To prove this point, we have computed the probability distribution of the energy states
for $n=317$ charges confined in the disk geometry with three different types of initialization.
In all cases the outer shell occupation has been fixed to $n_1^H=126$
predicted by Eq.\,(\ref{Scaln1}). In fact, this value corresponds to
the actual value associated with the MD absolute energy minimum.

As it is demonstrated in Ref.\,\onlinecite{toni},
there is a remarkable agreement between the CM and MD occupations even
for a charge number of subsequent shells.
Hence, we aim to assess the effectiveness of the CM prediction
for $n_2^H$, Eq.(\ref{n2}), as a guide to
initialize the external particle positions in the MD. To this end we have considered initial
configurations characterized by external occupations: $n_1=126$ (Set 1); $n_1=126,\, n_2=55$ (Set 2);
$n_1=126,\, n_2=56$ (Set 3).
We have generated 3650, 2000 and 2000 configurations, respectively.
In each case $n_1$ particles were initially set on the boundary at $r_1=1$,
and for the last two sets $n_2$ particles have been
homogeneously distributed at $r_2=0.96$.

In order to preserve these external shell occupations the $n_1, (n_2)$,
particles are frozen at a first stage, until
the inner particles slow down (typically after some $500$ time steps).
At a second stage, all particles are taken
into account and evolved according to Eq.\,(\ref{EqMDad}).
It is worth noticing that
the chosen value $r_2=0.96$ is higher than the CM result ($r_2\simeq 0.91$).
The reason is twofold: i) it guarantees the
desired $n_2$ value for the equilibrated
final configuration; ii) at the beginning of the second stage it provides  additional
excitation energy in the form of monopole oscillations that help to access low energy states.
The remaining particles are initialized randomly in the central region.

\begin{figure}[bth]
\vspace{-0.6cm}
\hspace{-1.5cm}
\includegraphics[scale=.55]{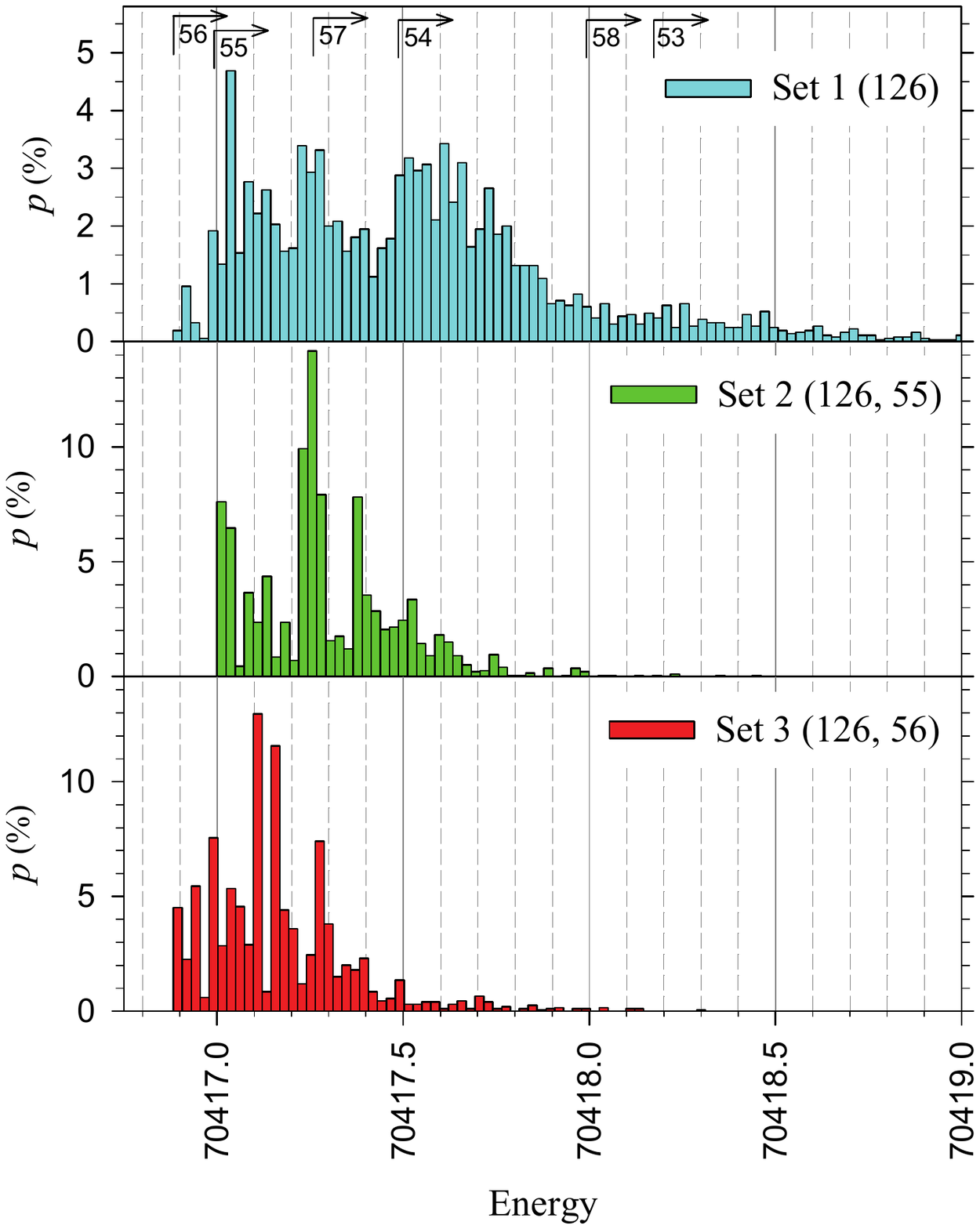}
\vspace{-6cm}
 \caption{(Color online)
  Histograms for energy states of $n=317$ charges confined in the disk geometry.
  The results are obtained
  with the aid of the MD method by using different initialization procedures. See text for details.}
 \label{Histog}
 \end{figure}

The results for the Set 1 (see Fig.\,\ref{Histog}, top panel)
consists of different final configurations with
$n_2=53 (5\%), 54 (36\%), 55 (38\%), 56 (18\%), 57 (3\%)$ and $58 (0.2\%)$.
The arrows indicate the starting energy for each $n_2$ value.
In this case the low energy states are dominated by configurations with
$n_2=(55,\, 56)$, although just about
one out of five runs leads to the $n_2$ value shared by the ground state at
${\cal E}_{\rm MD}=70416.883$.
The realization of the ground state does not exceed $0.2\%$.
The Set 2 (Fig.\,\ref{Histog}, middle panel) provides
the lowest state ${\cal E}_{\rm MD}=70417.000$
with $n_2=55$ (the CM partition), which is slightly above the true ground state.
The Set 3 (Fig.\,\ref{Histog}, bottom panel) explores the
nearby $n_2=56$ configurations, with $n_2=n_2^H$ provided by Eq.(\ref{n2}).
In this case the absolute energy minimum is found
with a probability $\sim 4.5\%$ which is higher by a factor 25 relative
to the probability found for the Set 1. Thus, even if one has to check nearby $n_2^H \pm 1$ values, the scanning effort clearly benefits from the
scalings found within the CM.

\begin{figure}[bth]
\vspace{-1.2cm}
\hspace{-0.91cm}
\includegraphics[scale=.232]{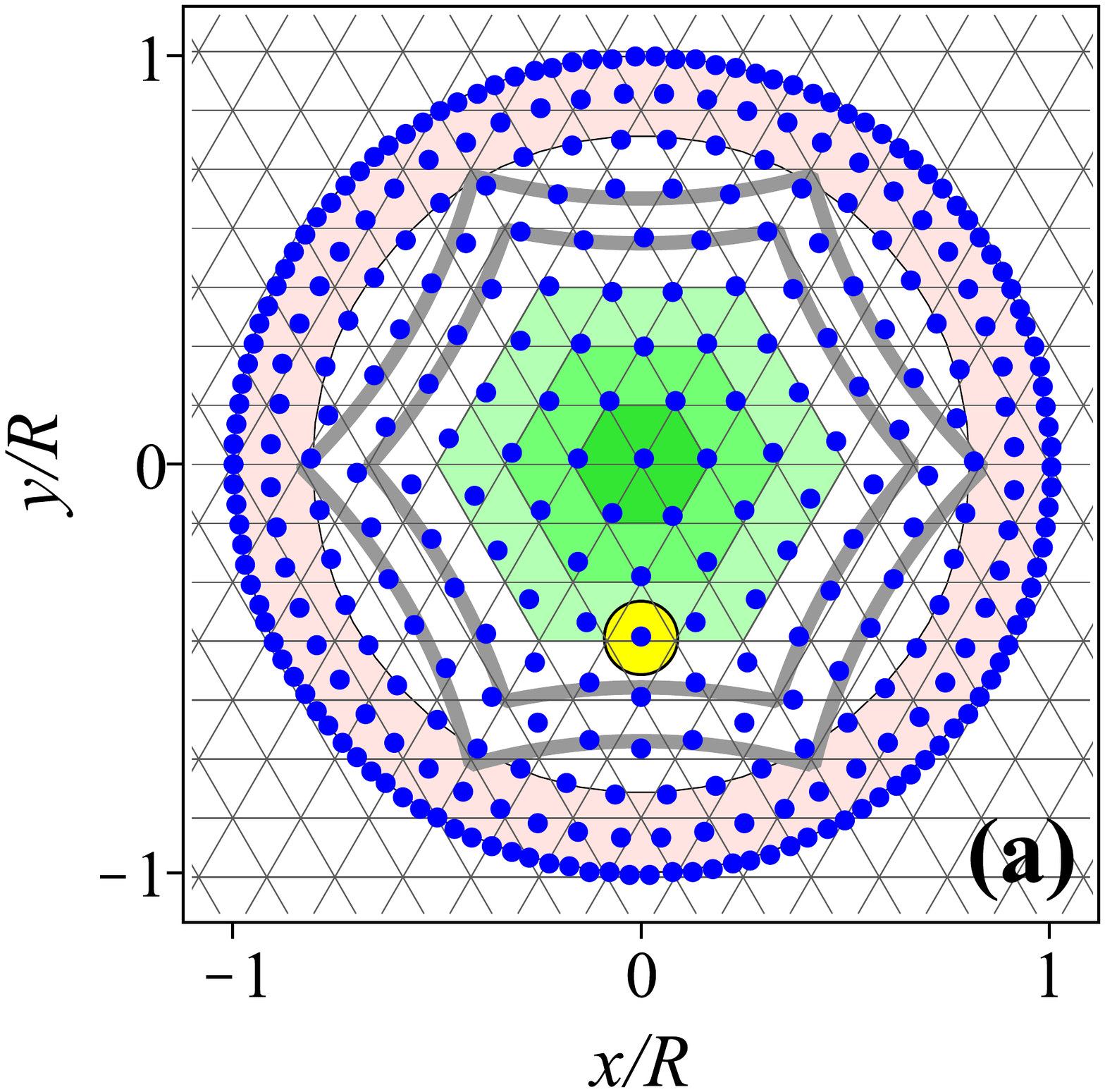}
\hspace{-.45cm}
\includegraphics[scale=.23]{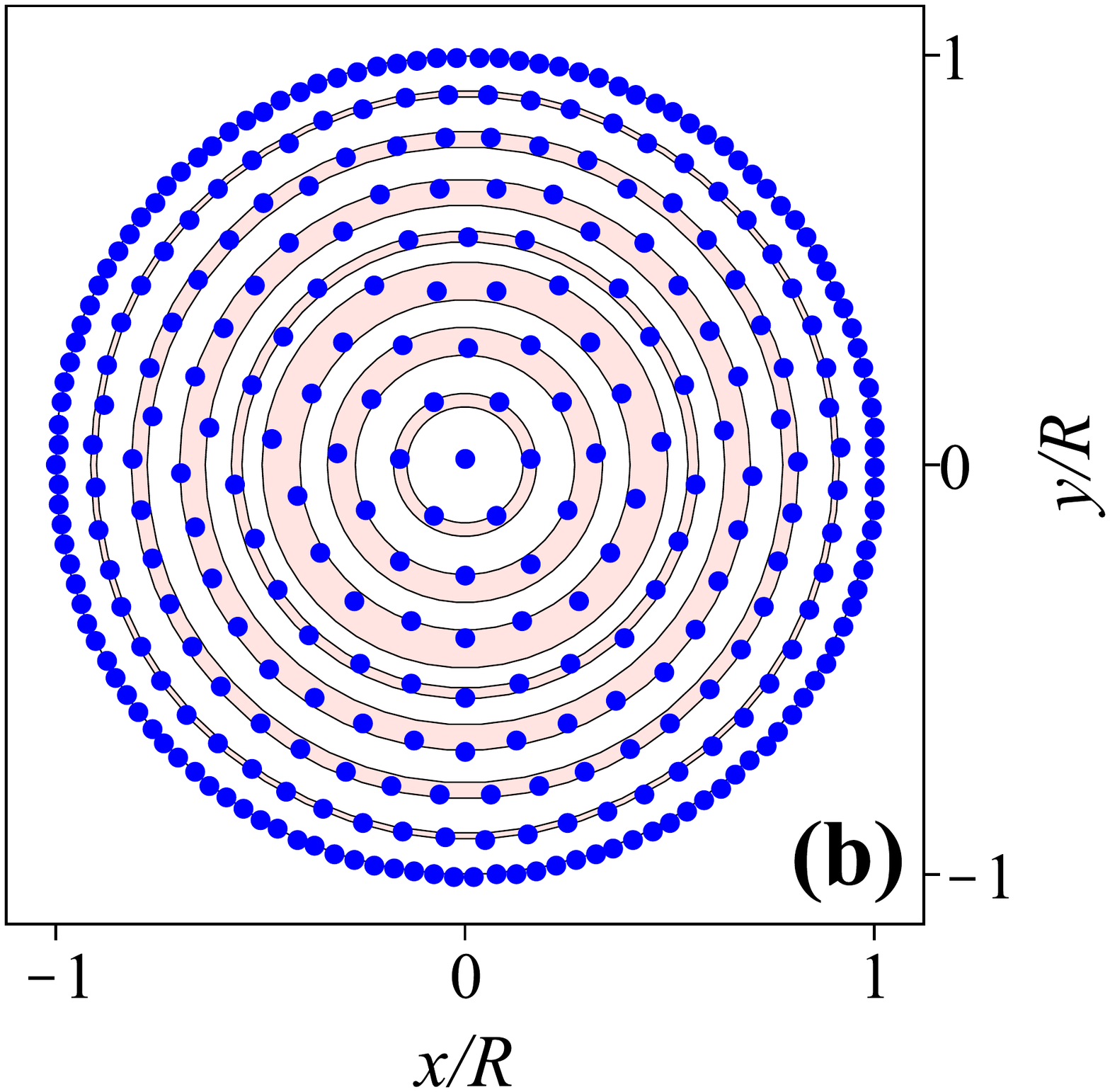}
\vspace{-1.2cm}
 \caption{(Color online)
  Structure of equilibrium configurations for the disk geometry obtained by
  means of the MD for $n=317$. (a)  Distribution of particles {\em versus} hexagonal grid.  
  The {\it core} (green) region with ${\{\bf 1}, {\bf 6}, {\bf 12}$) particles 
  exhibits a hexagonal pattern. The additional particle 
  (displayed within the small yellow circle) in the third shell breaks the hexagonal $6\,p$ rule.
 The intermediate region can still be associated with a deformed hexagonal lattice (gray lines to guide the eye). The {\it valence} shells contain $126,56,42$ charges with 
 an almost perfect circular structure. (b) Distribution of particles on various rings, 
 found with the clustering algorithm described in Sec.\, IV.C, with number of charges: 126,55,42,32,25,18,12,6,1.  
   }
 \label{def}
 \end{figure}

For the sake of illustration, we present in Fig.\,\ref{def} the distribution of charges 
in the equilibrium configuration obtained by means of our procedure for $n=317$.
Three regions are found (see Fig.\,\ref{def}a). The central region (green colored hexagonal area)
is comprised of the almost perfect CHL with 1,6 and 12 particles 
followed by a third shell containing 19 (instead of 18) particles.
This additional charge (at the center of the small yellow circle) breaks 
 the hexagonal structure. Its effect propagates to the middle region.
 Here, together with additional defects, it builds still a hexagonal, although 
 deformed, structure (gray lines are used to indicate the deformed lattice).
 The last region contains three external rings with 
126,56,42 particles.
Applying a simple clustering algorithm, 
that will be discussed below, we can identify circular shells in the MD results 
(see Fig.\,\ref{def}b). As a result, we obtain the following configuration  
(126,56,42,33,22,19,12,6,1) compatible with 
the CM result (126,55,42,32,25,18,12,6,1) (see also \cite{sup}).

Below, we compare the MD and the CM results in more details.

\subsection{Molecular dynamics and the circular model}

The numerical solution of the system (\ref{Ecoll}),\!\! (\ref{req1}) for the hard confinement
%\!\! (\ref{paren}), \!\! (\ref{req}),
provides a remarkable agreement
with the MD calculations for equilibrium configurations up to $n\!=\!105$,
excluding a few cases (see Table I in Ref.\,\onlinecite{cnp}).
Our MD results agree with those of Ref.\,\onlinecite{wor} up to
$n\!=\!160$ particles, while we obtain lower energies for $n\!=\!400,500,1000$ and also systematically better
values for $n>52$ than those implied in Fig. 8 of Ref.\,\onlinecite{mug}.
In Ref.\,\onlinecite{toni} the reader  can find the comparison of the MD and the CM results for the
disk geometry for $161\leq n\leq260$. In that paper, we have also demonstrated the usefulness of the CM to speed up the random search method of the true ground state for $n=395$ in the MD calculations, found in Ref.\cite{am}.

\begin{table}[htb]
\caption{Values for the only eleven cases where optimal configurations, obtained with
the aid of Eq.\,(\ref{req}), disagree with
the MD results. The systematic MD results for $n\leq 52$ can be found also in Ref.\cite{pet1}.
}
\begin{tabular}{|c|c|c|c|c|}
\hline
$n$  & ${{\cal E}^S}_{\rm avg}(n)$ & $\delta$ &  {\rm configuration} \\
\hline
  19  & 115.1127   &  0.0208  & $(11,7,1)_2^1$   \\
  22  & 149.7743   &  0.0571  & $(13, 8, 1)_1^3$   \\
  32  & 290.7905   &  0.0655  & $(16, 11, 5, 0)_1^4$  \\
  34  & 323.4146   &  0.0537  & $(16, 11, 6, 1)_1^2$ \\
  36  & 357.5053   &  0.0313  & $(16, 12, 7, 1)_3^1$ \\
  39  & 411.3570   &  0.0392  & $(18, 13, 7, 1)_1^4$ \\
  40  & 430.0216   &  0.0924  & $(18, 13, 8, 1)_1^4$ \\
  41  & 449.0085   &  0.1290  & $(18, 13, 8, 2)_1^2$ \\
  46  & 548.6758   &  0.0976  & $(18, 15, 9, 4)_4^1$  \\
  52  & 678.9715   &  0.0671  & $(20, 16, 11, 5, 0)_{1,2}^{4,5}$  \\
  53  & 701.8207   &  0.1162  & $(19, 16, 11, 6, 1)_{1}^{2}$    \\

\hline
\end{tabular}
\label{table:tb1}
\end{table}
In the case of the parabolic potential we obtain good agreement with the MD results
as well, excluding a few cases (see Table\,\ref{table:tb1}) up to $n\approx 51$.
The difference $\delta= {{\cal E}^S}_{\rm avg}-{\cal  E}_{\rm MD}$  provides the error of our approximation.
The rings are counted starting from the external one
which is the first ring. The notation $(11, 7, 1)_2^1$  means that we have to add
one particle in the first ring and remove one particle from the second ring in order
to obtain the MD result. Although the total energy errors are very small, the assumptions of our
model fail to predict the correct configurations for shown total $n$.

Since in the disk geometry the external radius is fixed ($R=1$),  the parabolic potential
has one more degree of freedom in terms of collective variables. It is natural to assume
that this degree of freedom is related to fluctuations of the external ring radius
around some equilibrium radius value (radial fluctuations).
As a result, such a motion creates fluctuations in the particle number
around some optimal value in the external ring affecting
the particle number in the closest ring.

In order to get deeper insight into this phenomenon we have applied a simple
clustering algorithm to identify the formation of circular shells in the MD results.
At a first stage, we order particles according
to their distance to the centre $r_1\leq r_2 \leq \cdots \leq r_n$.
Next, we define the gaps between consecutive particles $\delta_i =r_{i+1}-r_i$
and sort them by decreasing value, {\em i.e.},
$\delta_{i_1} \geq \delta_{i_2} \geq \cdots \geq \delta_{i_{n-1}}$.
By defining the function
\beq
\label{EqFMD}
F_{\rm MD} (p)=\frac{r_1+\sum_{k=1}^{p-1} \delta_{i_k} } {p}\, ,
\eeq
the optimal groups are found by maximizing the average spatial separation between
consecutive groups with respect to the number of shells $p$. Additionally, we impose
the constraint $0\leq p(n)-p(n-1)\leq 1$ and detect when a new particle settles
at the centre of the structure, $r_1\simeq 0$, thus opening a new shell.
Once the number of shells is obtained, the related function
\beq
\label{EqRMD}
R_{\rm MD}(n)=\frac{p\, F_{\rm MD}(p)}{r_n}
\eeq
provides a simple measure of how close the MD particle configuration is to a well
defined ring structure.  Notice that for strict circular configurations, such as those provided
by our circular model, $R_{\rm MD}(n)=1$. A significant departure from this maximum
value would imply that the particle distribution deviates from the prediction of the CM.

\begin{figure}[ht]
\vspace{-1.4cm}
\hspace{-0.91cm}
\includegraphics[scale=.232]{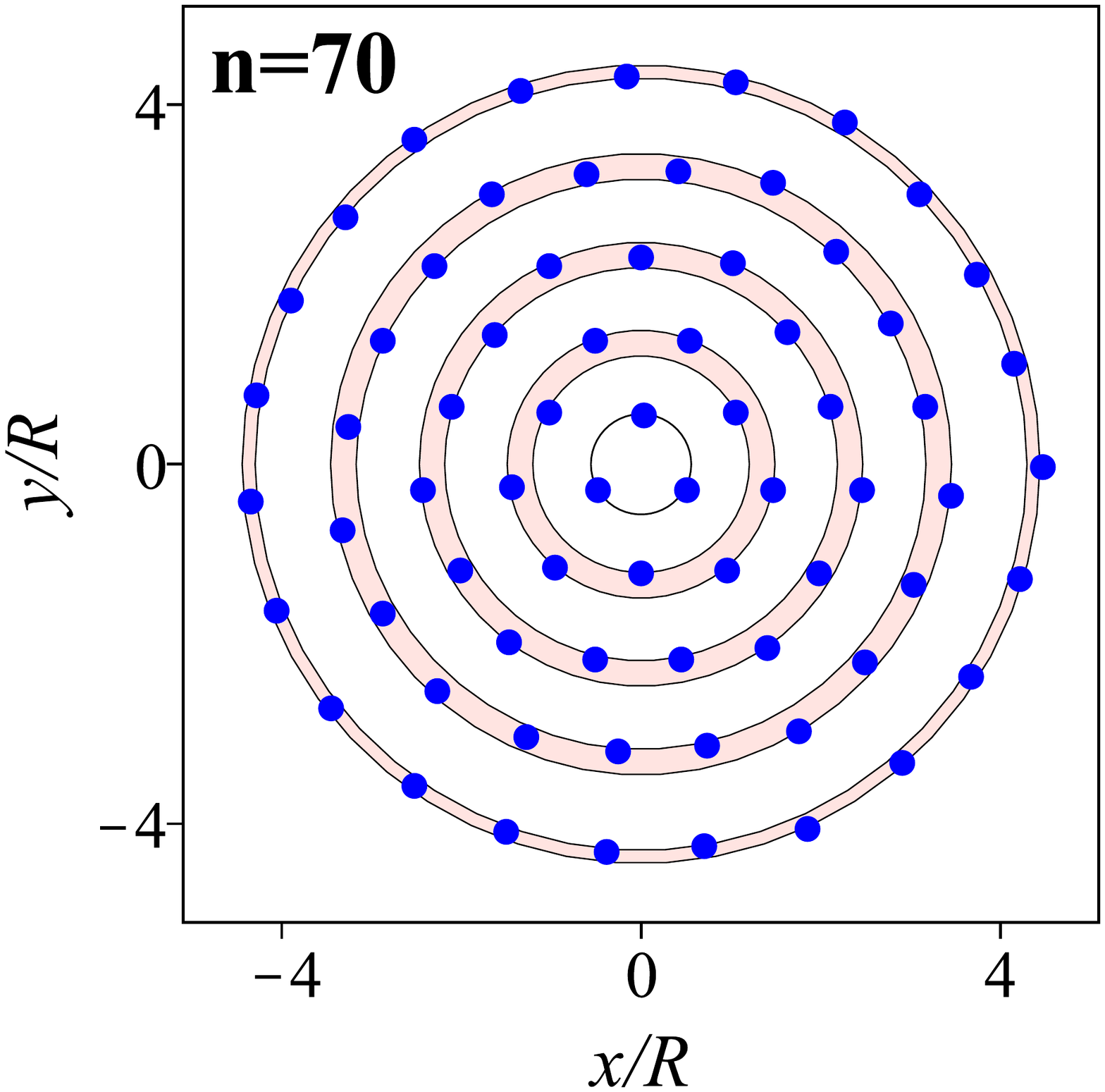}
\hspace{-.45cm}
\includegraphics[scale=.23]{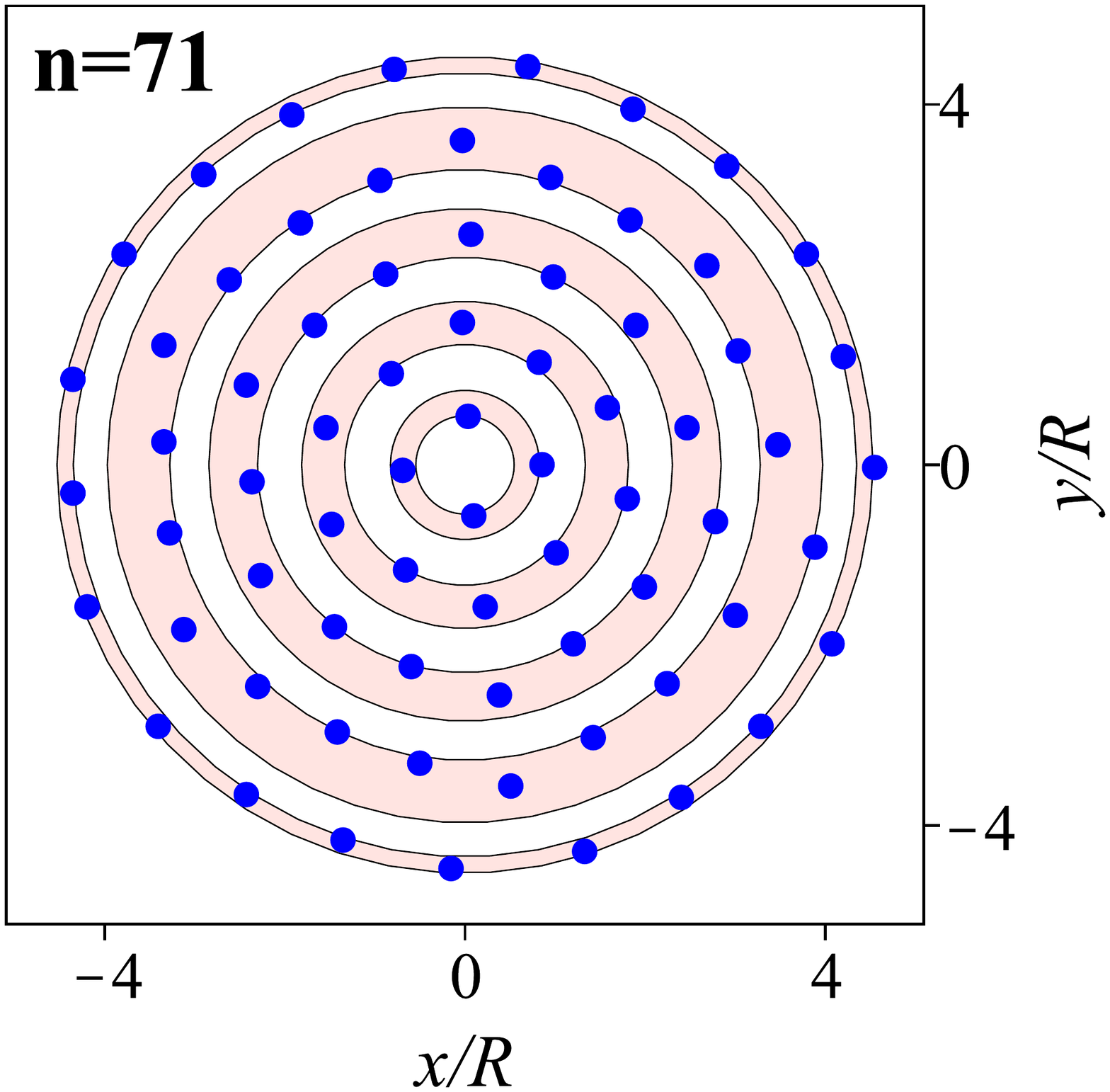}
\vspace{-1.2cm}
 \caption{(Color online)
  Structure of the MD equilibrium configurations for $n=70,71$ charged particles in
   the circular parabolic potential.
  In both systems five rings are identified. For $n=70$ the MD
  configuration is $ {\{\bf 3}, {\bf 9}, {\bf 15}, {\bf 20}, {\bf 23}\}$,
  ${\cal  E}_{\rm MD}=1135.298$. The circular model predicts the same configuration
  with ${\cal E}_{\rm avg}^S=1135.474$.
  For $n=71$ the MD (the circular model)
  configuration is
  $ {\{\bf 4}, {\bf 10}, {\bf 15}, {\bf 21}, {\bf 21}\}$
  $({\{\bf 4}, {\bf 9}, {\bf 15}, {\bf 20}, {\bf 23}\})$,
  ${\cal  E}_{\rm MD}({\cal E}_{\rm avg}^S)=1163.410 \, (1163.579)$. }
 \label{MDss}
 \end{figure}

\begin{figure}
%\centering
\hspace{-1cm}
\includegraphics[scale=0.45]{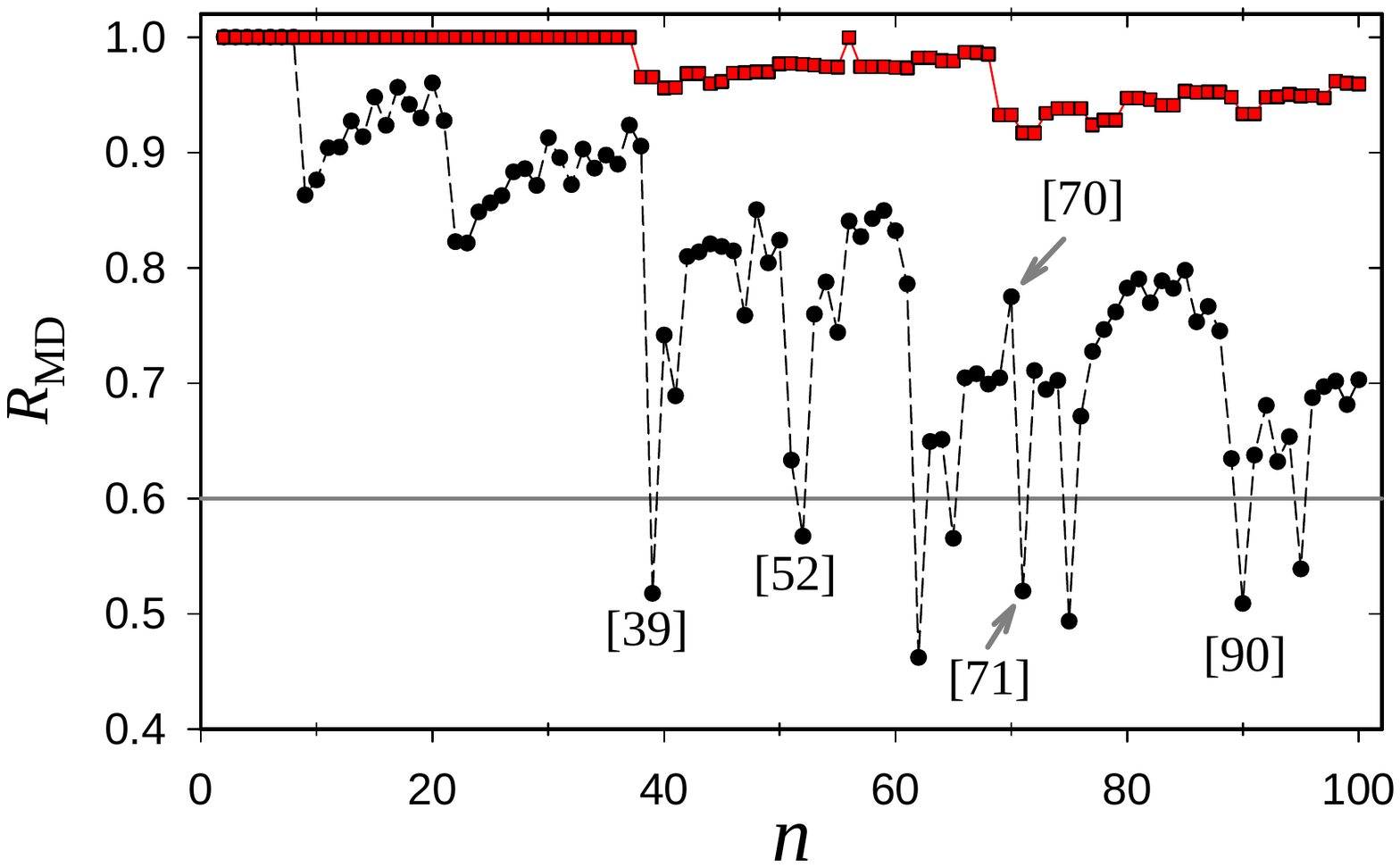}
\vspace{-8.cm}
  \caption{(Color online)
  The function (\ref{EqRMD}) {\em vs} number of charged particles.
  The results for the disk geometry and the parabolic potential
  are shown by (red) squares and (black)  circles. The arrows indicate the location
  of $n=70,71$ charge particle configurations discussed on Fig.\ref{MDss}.}
  \label{F3}
\end{figure}

This fact is illustrated in
Fig.\,\ref{MDss} where the result of this clustering algorithm is shown for $n=70,71$.
The actual values $R_{\rm MD}(70)=0.775$ and
$R_{\rm MD}(71)=0.520$ indicate that the system with
$n=70$ is quite well described by a ring structure. It is
not the case, however, for $n=71$, where the resulting shells have much larger widths.
The addition of one particle produces a visible finite size effect which transforms
the system  from a rigid ring organization to a kind of glasslike behavior.
In Fig.\,\ref{F3} the function (\ref{EqRMD}) shows that the MD configurations
form a robust circular structure for the disk geometry; its value remains
close to the maximum $(\geq 90\%)$ for $n\leq 100$ charged particles.
The accumulation of a big fraction of particles on the perfect circular boundary
strongly constrains the internal ring organization.
In this case the circular model provides a remarkable agreement with the MD
results both for the equilibrium configurations and their energies.

Conversely, while this function does frequently not drop down below
$(\simeq 60\%)$ in the case of the parabolic potential,
there are a few cases $n=39,52,62,65,71,75,90...$ that exhibit a visible deviation
from a circular distribution.
In these cases the system manifests detectable fluctuations in
particle positions, driven by a change in the central structure
($n_p\!=\!1 \to 2 \ {\rm or} \ n_p\!>\!1 \to 1$), that
affect the width of the corresponding shells.
In general, one observes a kind of cold melting of the system
configuration that preserves, however, to some extent the ring structure.

A comparison of the MD results with those of the circular model
for the total energy  demonstrates a remarkable agreement.
Although there is a small disagreement between the results obtained
within our model and MD calculations for the ground state configurations,
it is noteworthy that the {\sl onset of the centered
hexagonal lattice} in the MD calculations for large $n$ has been  clearly recognized
with the aid of the circular model.

\section{Summary}
We have developed the method (see Sects.\,II,\,III) which allows us
to analyze the equilibrium formation and filling of rings with a finite number of particles
interacting by means of the Coulomb forces in the case of a circular lateral confinement.
As an example, we analyzed  disk and parabolic confinements.
Our approach is based on the cyclic symmetry
of the Coulomb energy between particles distributed over different rings. As a result,
the problem of $n$ interacting charged particles is reduced to
the description of $p\,(\ll\!\! n)$ rings,
with homogeneously distributed integer charges.
To test the validity of the method we have also developed the MD approach
(Sec.\,IV) and present the comparison of the results in Ref.\,\onlinecite{sup}.

We have demonstrated that  our method is good enough
to obtain exact ground state configurations with correct energies,
excluding a few particular cases, up to $n\leq\!105$  and $\leq\!51$ for a hard
and parabolic confinements, respectively.
For bigger systems the solution of the model equations provide also
very good approximations to the exact ground state configurations.
 Indeed, the energy errors do not exceed a small percentage fraction of the exact values.
 However, this achievement is a necessary but not a sufficient condition
 to conclude that the CM is effective.  The systematic analysis of the CM results
provides the estimate for a number of charges $n_1$ distributed over
the external (boundary) ring for large systems $(n\le400)$.
Once this number is identified, one has to fix the number of charges on
the second ring, the nearest neighbor to the boundary one, based on
the CM prediction $n_2\pm1$. At the same time, the inner shell structure,
as predicted by the CM, should be disregaded, since it may introduce a bias.
With the aid of this strategy
the computational effort to get global energy minima is much less than
in the MD or simulated annealing (SA) calculations.
In fact, both methods are flawed by two major problems.
First, due to the long range of Coulomb forces, computing time grows with the number of
particles as $n^3$. Second, and more important, the number of equilibrium
configurations near the absolute energy minimum grows exponentially with $n$. This last fact
 increases considerably the computational efforts needed to avoid getting stuck
 at local energy minima, because many different initial condition simulations are needed.
Therefore, we believe that an accurate method, capable of explaining shell structure,
which works with a much reduced number of variables ($\sqrt{n/2}$ {\em vs} $n$)
is a remarkable achievement.
Moreover, the results obtained by means of our method can be used to {\it feed} SA or MD
calculations with sensible initial configurations, reducing substantially the amount of scanning
normally needed to {\it visit} the global energy minima.

In our circular model each $p\!-\!{\rm th}$ shell consists of
a set of point charges distributed over
a regular hexagon inscribed and circumscribed by two circles.
Based on the results of our analysis, we have found that in both potentials the increase of particle
number leads to the onset of a {\sl centered hexagonal lattice} that transforms smoothly to
a few circular rings at the boundary. A similar conclusion has been drawn
on the basis of the Monte Carlo calculations for $n>150$ charges
confined by a circular parabolic potential \cite{pet3},
although the authors  admitted that their equilibrium
configurations are not necessarily true ground states. Based on our results
we speculate that this self-organization should be typical for any 2D finite system of
identical charges confined by a circularly symmetric potential.
We recall, however, that depending on the size of the system one has to take into account
 the onset of quantum correlations for increasing particle number at a fixed system size
(see a discussion in  Ref.\,\onlinecite{cnp}).

\section*{Acknowledgments}
We thank Kostya Pichugin for useful discussions, and
F.M. Peeters for bringing to our attention Ref.\onlinecite{pet3}.
M.C. is grateful for the warm
hospitality at JINR. This work was supported in
part by Bogoliubov-Infeld program of BLTP and
Russian Foundation for Basic Research.

\appendix
\section{Cyclic symmetry and the Coulomb sums}
\label{red}
We want to prove that
\[ \sum_{i=1}^n \sum_{j=1}^m \,\textrm{F}\!\left \lbrack \cos\left ( 2\pi\left( \frac{i}{n}-
\frac{j}{m}\right) \right) \right \rbrack=
   G\,\sum_{k=1}^L\, \textrm{F}\!\left \lbrack \cos\left ( 2\pi \frac{k}{L} \right) \right \rbrack\ , \]
where $L\equiv {\rm LCM}(n,m)$ and
$G\equiv{\rm GCD}(n,m)=n\times m/L$ are the  least common multiple and
greatest common divisor of the  of numbers $(n,m)$, respectively.

\bigskip
{\bf Proof}:

\smallskip
Due to the fact that $\textrm{F}$ is a function (every element belonging to the domain is related
 to a unique element of the image set) it suffices to prove that the multiset \footnote{A multiset
 is a set which can contain repeated elements.} of angles
 $A=\Bigl \lbrace 2\pi \left(\frac{i}{n}-\frac{j}{m} \right),\;i=1,\ldots, n \,\wedge\,
 j=1,\ldots, m \Bigr \rbrace$
 is equal, under the cyclic symmetry, to the multiset $B=\Bigl \lbrace 2\pi\frac{k}{L}\,, 
 k=1\ldots L \Bigr \rbrace$, where each element is repeated $G$ times.

\smallskip
In order to take into account the cyclic symmetry explicitly, we change the numbers $k$ by their
associated equivalence classes, defined as
$C_k=\Bigl\lbrace k+\gamma\mspace{1mu} L\,,\:\forall\, \gamma\in\mathbb{Z}\Bigr \rbrace$.
As a result, a $k$-value appearing inside the $B$ multiset has to be read as an unspecified member
of the $C_k$ class. The set composed by all these classes constitutes a partition of
$\mathbb{Z}$ into $L$ classes.

Since any linear combination of two integers $(n,m)$ is equal to a multiple of its greatest common
divisor $G={\rm GCD}(n,m)$, we have
\begin{equation}\label{ec1}
\frac{i}{n}-\frac{j}{m} = \frac{i\mspace{1mu}m-j\mspace{1mu}n}{n\mspace{1mu}m}=\frac{k\mspace{1mu}G}{n\mspace{1mu}m}=\frac{k\mspace{1mu}G}{L\mspace{1mu}G}=\frac{k}{L}\,,
\end{equation}
where $k$ is an integer.
Evidently, this result allows us to change two sums over the variables $i$ and $j$
by a single suitable sum over the variable $k$.

Note that for the multiset of angles $A$
%due to the fact that the angles are the argument of a periodic function,
the substitutions
\begin{align*}
    i \rightarrow i + \alpha\mspace{1mu}n & & \text{and} & & j \rightarrow j + \beta\mspace{1mu}m\ ,
\end{align*}
where $\alpha,\beta\in \mathbb{Z}$, have no practical effect. Furthermore, since a pair of indices $(i,j)$
produces an index $k$ that belongs to the class $C_k$, the shifted pair of indices
$(i+\alpha\mspace{1mu}n,j+\beta\mspace{1mu}m)$ produces
an index $k'$ that also belongs to the class $C_k$:
\bea
&    (i+\alpha\mspace{1mu}n)\mspace{1mu}m - (j+\beta\mspace{1mu}m)\mspace{1mu}n =
    i\mspace{1mu}m - j\mspace{1mu}n + (\alpha-\beta)\mspace{1mu}L\mspace{1mu}G =\nonumber\\
&    (k+(\alpha-\beta)\mspace{1mu}L)\mspace{1mu}G\,.\nonumber
\eea
Evidently, there are exactly $L$ different $C_k$ classes.
\smallskip

Now we show that each class contains G different elements.
If there exists a pair $(i,j)$ such that
\begin{equation}\label{ec2}
    i\mspace{1mu}m-j\mspace{1mu}n=k\mspace{1mu}G\,,
\end{equation}
then there are exactly $G$ pairs $(i',j')$, with the restrictions $1\le i' \le n$ and $1\le j' \le m$, such that $i'\mspace{1mu}m-j'\mspace{1mu}n=k\mspace{1mu}G$.
Note that if $(i,j)$ satisfy~(\ref{ec2}), then all couples of the form
\[ \left( i'=i + l\mspace{1mu}\frac{n}{G}\: , \, j'=j + l\mspace{1mu}\frac{m}{G}\right) \;,\]
where $l$ is an arbitrary integer, also satisfy ~(\ref{ec2}), because
\[ i'\mspace{1mu}m-j'\mspace{1mu}n = \left(i+l\mspace{1mu}\frac{n}{G}\right)m-\left(j+l\mspace{1mu}\frac{m}{G}\right)n =
 i\mspace{1mu}m - j\mspace{1mu}n = k\mspace{1mu}G\;.\]
If we restrict the values of $l$ to $0\le l \le G-1$, i.e. $G$ different choices, the resulting values
$i'$ and $j'$ are all different. Indeed, in this situation, if $G>1$, the following inequalities hold
\begin{align*}
    \frac{n}{G}\le |i'-i''|<n & & \text{and} & & \frac{m}{G}\le |j'-j''|<m\ .
\end{align*}
If $l=G+\lambda$, with $\lambda=0,1,2,\ldots$ the indices are cyclicly equivalent
\[ i+\frac{(G+\lambda)\mspace{1mu}n}{G} = i + n + \frac{\lambda\mspace{1mu}n}{G} \sim i +
\frac{\lambda\mspace{1mu}n}{G}\;. \]
Therefore, since there are $n\mspace{1mu}m$ pairs, every one of the $L$ different $C_k$ classes must
contain $G$ pairs. This completes the proof. The fact that we consider a \textrm{cosine}
function is irrelevant, since the requirement that $\epsilon (r_1,r_2,\theta)$ to be periodic is enough.

\section{Coefficients of the Fourier transform}
\label{coef}
In order to find an analytical expression for the integral (\ref{clm}), we employ
the Legendre expansion
for the Coulomb potential
\beq
\frac{1}{\left[r_1^2+r_2^2-2\,r_1\,r_2\,\cos \psi\right]^{1/2}}=
%\epsilon (r_1,r_2,\psi)=
\frac{1}{r_>}\sum_{u=0}^\infty t^u P_u(\cos \psi)\nonumber\, ,
\eeq
where
$r_>=\max(r_1,r_2)$, $r_<=\min(r_1,r_2)$, $t=r_</r_>$.
With the aid of the cosine series for the Legendre polynomials  (see Ref.\,\onlinecite{AS}, p.\,776)
\begin{equation}
\label{LegendreCos}
    P_u(\cos \psi) = \frac{1}{4^u}\sum_{v=0}^u \binom{2v}{v} \binom{2(u-v)}{u-v} \cos\left( (u-2v)\psi\right)\nonumber\, ,
\end{equation}
we obtain for Eq.~(\ref{clm}) the following form
\bea
\label{exprs2}
  C_{\ell nm}(r_1,r_2)&=&\frac{n m}{r_>\pi} \sum_{v=0}^\infty \left(\frac{t}{4}\right)^v \binom{2v}{v}
  \sum_{w=0}^v \left(\frac{t}{4}\right)^w \binom{2w}{w}\nonumber\\
  &\times&\int_0^{2\pi} \cos\left[(w-v)\psi\right] \cos(\ell L \psi) \,d\psi\,,
  \eea
where $u=w+v$. The integral splits in two terms, since
$\cos A \cos B = \left[ \cos(A+B) +\cos(A-B) \right]/2$.
In order to have a nonzero value, the first term with the argument $w=v-lL$
in the cosine function  requires $v\geq l L$.
 As a result, we have
\beq
 C_{\ell nm}(r_1,r_2)=\frac{n m}{r_>} \sum_{v=\ell L}^\infty \left(\frac{t}{4}\right)^{2v-\ell L}
 \binom{2v}{v} \binom{2(v-\ell L)}{v-\ell L}\nonumber
\eeq
\beq
\label{exprs6}
=\frac{n m}{r_>} \left(\frac{t}{4}\right)^{\ell L}\sum_{k=0}^\infty \binom{2(\ell L+k)}{\ell L+k}
    \binom{2k}{k}\, \left(\frac{t}{4}\right)^{2k}\,,
\eeq
where $k=v-\ell L$. The second term yields the same result, that
duplicates the expression (\ref{exprs6}).

This double sum can be expressed in terms of the hypergeometric function.
In virtue of the identity
\beq
\frac{1}{4^p}=\frac{\Gamma(p+1/2)}{\sqrt{\pi}{p!}}\nonumber\, ,
\eeq
we arrive at the final form
\bea
&C&\!\!\!\!_{\ell nm}(r_1,r_2)=\nonumber\\
&=&\frac{2 n m}{r_>}\, \frac{t^{\ell L}}{\pi}\,\sum_{k=0}^\infty
\frac{\Gamma(\ell L+k+1/2)\Gamma(k+1/2)}{\Gamma(\ell L+k+1)k!}  t^{2k}\nonumber\\
\label{ff}
&=&\frac{2 n m}{r_>}\, \frac{t^{\ell L}}{\sqrt{\pi}}\,\frac{\Gamma(\ell L+1/2)}{(\ell L)!}\, \sum_{k=0}^\infty \frac{(1/2)_k(\ell L+1/2)_k}{(\ell L+1)_k} \frac{t^{2k}}{k!}\nonumber\\
&=&\frac{2 n m}{r_>}\, \frac{(2\ell L-1)!!}{(\ell L)!}\left( \frac{t}{2} \right)^{\ell L}
   \times\nonumber\\
    &&\times {}_2 F_{\mspace{-1mu}1} \Bigl( 1/2 \,,\ell L+1/2 \,;\ell L +1 \,;\, t^2\Bigr)\,.
\label{fincef}
\eea
Here, we have used a definition of the hypergeometric function in terms of
the Pochhammer symbol $(a)_b=\Gamma(a+b)/\Gamma(a)$.

At $\ell L\gg1$  one can use the asymptotic value for the central  binomial coefficients
$\binom{2M}{M}/(4^M) \approx 1/\sqrt{\pi M}$ ,
where $M=\ell L$. As a result, we obtain the asymptotic limit of Eq.\,(\ref{fincef}):
\begin{equation}
\label{exprs7}
    C_{\ell nm}(r_1,r_2)\approx\frac{2 n m}{\sqrt{\pi}\,r_>} \, \sum_{k=0}^\infty
    \frac{\binom{2k}{k}}{4^k} \frac{t^{\ell L+2k}}{\sqrt{\ell L+k}}\ .\nonumber
\end{equation}
This result shows explicitly the decreasing magnitude of
the coefficients accompanying the powers of the variable $t$ in Eq.\,(\ref{ff1})
with the increase of the product $\ell\, L$.

\newpage
%\renewcommand{\thesection}{\arabic{section}}
%\section
\begin{widetext}
\begin{center}
{\bf \large Supplemental Material: The MD and Circular Model results}
\end{center}
 \vspace{1cm}
\centerline{\bf I: Disk geometry}
 \vspace{1cm}
These tables summarize our results corresponding to
 the minimum energy equilibrium configurations under disk confinement,
 discussed in Sec.IIIA.
  \vspace{1cm}
 \\
 \\
\centerline{DATA for Table (Eq. (18))}
\\
\\
%\begin{table}[t]
%\begin{center}
\hspace{-0.35cm}
\begin{tabular}{c llll}
\hline\hline
n  & CM energies & Configuration & MD energies & Configuration\\ [0.5ex]
\hline
11:  & 48.57568  &  [11]                           & 48.57568    & [11]                         \\
29:  & 444.5491  &  [23,6]                         & 444.5478    & [23,6]                       \\
55:  & 1792.007  &  [37,13,5]                      & 1791.974    & [37,13,5]                    \\
90:  & 5115.563  &  [53,20,12,5]                   & 5115.408    & [52,20,12,5,1]                \\
135: & 11995.371 &  [70,29,19,12,5]                & 11994.978   & [70,28,19,12,5,1]             \\
186: & 23391.044 &  [87,37,26,19,12,5]             & 23390.284   & [87,37,26,18,11,6,1]          \\
246: & 41743.132 &  [106,46,34,25,18,12,5]         & 41741.995   & [106,46,34,25,16,12,6,1]      \\
316: & 69962.348 &  [126,56,42,32,25,18,11,5]      & 69960.435   & [126,55,42,33,26,15,12,6,1]    \\
394: & 110093.60 &  [147,66,50,40,32,25,18,11,5]   & 110090.41   & [147,66,50,40,26,26,19,13,6,1] \\
\hline
\end{tabular}
\vspace{2cm}
 \\
 \\
 \centerline{DATA for Table (Eq. (20))}
\\
\\
\hspace{-1cm}
\begin{tabular}{c llll}
\hline\hline
n  & CM energies & Configuration & MD energies & Configuration\\ [0.5ex]
\hline
12:  & 59.57568  &  [11,1]                           & 59.57568    & [11,1]                         \\
30:  & 479.0854  &  [23,6,1]                         & 479.0796    & [23,6,1]                       \\
56:  & 1862.734  &  [37,12,6,1]                      & 1862.650    & [37,12,6,1]                    \\
92:  & 5358.578  &  [53,20,12,6,1]                   & 5358.353    & [53,20,12,6,1]                \\
136: & 12181.755 &  [70,28,19,12,6,1]                & 12181.345   & [70,28,19,12,6,1]             \\
187: & 23652.947 &  [87,37,26,18,12,6,1]             & 23652.188   & [87,37,26,18,12,6,1]          \\
248: & 42447.440 &  [106,46,34,25,18,12,6,1]         & 42446.278   & [107,46,34,25,17,12,6,1]      \\
317: & 70418.854 &  [126,55,42,32,25,18,12,6,1]      & 70416.883   & [126,56,42,33,22,19,12,6,1]    \\
395: & 110667.59 &  [147,65,50,40,32,24,18,12,6,1]   & 110664.44   & [147,66,51,40,26,26,19,13,6,1] \\
\hline
\end{tabular}
%\end{widetext}
%
\vspace{1cm}
\newpage
%\begin{widetext}
%\section{Parabolic confinement}
\centerline{\bf II: Parabolic confinement}
\vspace{1cm}
 These tables summarize our results corresponding to
 the minimum energy equilibrium configurations under parabolic confinement. $E_{\rm MD}$,
 $E_{\rm CM}$ are the total MD and the Circular Model energies. $R_{\rm avg} $ and $R_{\rm ext}$
 are respectively the average radius for the external shell in MD and the external CM radius. All quantities are expressed in parabolic units, defined in the main text. The approximate shell configuration
 in MD has been obtained by the algorithm described in Sec.IVB. Notice that for specific
 values of $n$ (e.g., $n=90$)  some intermediate shells appear grouped together,
 indicating rather big particle dispersions within those shells.
\begin{table}[h]
\centering
% \vskip 1 mm
% text moved outside
{\LARGE  Results for $2\le n \le 33$\par}
\begin{tabular} {@{\hspace{0.15cm}} c @{\hspace{0.5cm}} c @{\hspace{0.5cm}} r @{\hspace{0.5cm}} c @{\hspace{0.25cm}} | @{\hspace{0.25cm}} c @{\hspace{0.5cm}} r @{\hspace{0.5cm}} c @{\hspace{0.15cm}}}
\hline
$n$  & $E_{\rm MD}$ &  Configuration &  $R_{\rm avg}$  & $E_{\rm CM}$
 & Configuration & $R_{\rm ext}$\\
\hline
2 & 1.190551 & [2] & 0.630 & 1.190551 & [2] & 0.630\\
3 & 3.120126 & [3] & 0.833 & 3.120126 & [3] & 0.833\\
4 & 5.827177 & [4] & 0.985 & 5.827177 & [4] & 0.985\\
5 & 9.280127 & [5] & 1.112 & 9.280127 & [5] & 1.112\\
6 & 13.35587 & [5,1] & 1.334 & 13.35587 & [5,1] & 1.334\\
7 & 17.99543 & [6,1] & 1.414 & 17.99543 & [6,1] & 1.414\\
8 & 23.29609 & [7,1] & 1.490 & 23.29609 & [7,1] & 1.490\\
9 & 29.20266 & [7,2] & 1.640 & 29.24607 & [7,2] & 1.646\\
10 & 35.59701 & [8,2] & 1.699 & 35.63458 & [8,2] & 1.704\\
11 & 42.47199 & [8,3] & 1.830 & 42.50370 & [8,3] & 1.834\\
12 & 49.89776 & [9,3] & 1.877 & 49.93147 & [9,3] & 1.882\\
13 & 57.79314 & [9,4] & 1.992 & 57.81205 & [9,4] & 1.995\\
14 & 66.21506 & [10,4] & 2.033 & 66.23594 & [10,4] & 2.036\\
15 & 75.09498 & [10,5] & 2.134 & 75.10986 & [10,5] & 2.137\\
16 & 84.44850 & [10,5,1] & 2.228 & 84.49102 & [10,5,1] & 2.234\\
17 & 94.21966 & [10,6,1] & 2.320 & 94.23676 & [10,6,1] & 2.323\\
18 & 104.4085 & [11,6,1] & 2.349 & 104.4258 & [11,6,1] & 2.352\\
19 & 115.0919 & [12,6,1] & 2.377 & 115.1127 & [11,7,1] & 2.434\\
20 & 126.1922 & [12,7,1] & 2.460 & 126.2002 & [12,7,1] & 2.461\\
21 & 137.7733 & [13,7,1] & 2.486 & 137.7859 & [13,7,1] & 2.488\\
22 & 149.7172 & [12,8,2] & 2.608 & 149.7743 & [13,8,1] & 2.562\\
23 & 162.0612 & [13,8,2] & 2.632 & 162.1328 & [13,8,2] & 2.635\\
24 & 174.7901 & [13,8,3] & 2.701 & 174.8619 & [13,8,3] & 2.705\\
25 & 187.9243 & [13,9,3] & 2.768 & 187.9888 & [13,9,3] & 2.771\\
26 & 201.4657 & [14,9,3] & 2.788 & 201.5293 & [14,9,3] & 2.791\\
27 & 215.3890 & [14,9,4] & 2.852 & 215.4374 & [14,9,4] & 2.855\\
28 & 229.6947 & [14,10,4] & 2.913 & 229.7357 & [14,10,4] & 2.915\\
29 & 244.3994 & [15,10,4] & 2.931 & 244.4401 & [15,10,4] & 2.933\\
30 & 259.4762 & [15,10,5] & 2.990 & 259.5184 & [15,10,5] & 2.992\\
31 & 274.9301 & [15,11,5] & 3.046 & 274.9566 & [15,11,5] & 3.048\\
32 & 290.7250 & [15,11,5,1] & 3.101 & 290.7905 & [16,11,5] & 3.064\\
33 & 306.8574 & [15,11,6,1] & 3.156 & 306.9034 & [15,11,6,1] & 3.159\\
\hline
\end{tabular}
\end{table}
\vskip 5mm

\begin{table}
\centering
{\LARGE  Results for $34\le n \le 77$\par}
\begin{tabular} {@{\hspace{0.15cm}} c @{\hspace{0.5cm}} c @{\hspace{0.5cm}} r @{\hspace{0.5cm}} c @{\hspace{0.25cm}} | @{\hspace{0.25cm}} c @{\hspace{0.5cm}} r @{\hspace{0.5cm}} c @{\hspace{0.15cm}}}
\hline
$n$  & $E_{\rm MD}$ &  Configuration &  $R_{\rm avg}$  & $E_{\rm CM}$
 & Configuration & $R_{\rm ext}$\\
\hline
34 & 323.3609 & [15,12,6,1] & 3.206 & 323.4146 & [16,11,6,1] & 3.173\\
35 & 340.2139 & [16,12,6,1] & 3.221 & 340.2722 & [16,12,6,1] & 3.223\\
36 & 357.4740 & [17,12,6,1] & 3.235 & 357.5053 & [16,12,7,1] & 3.274\\
37 & 375.0712 & [17,12,7,1] & 3.286 & 375.0966 & [17,12,7,1] & 3.288\\
38 & 393.0063 & [17,13,7,1] & 3.334 & 393.0316 & [17,13,7,1] & 3.335\\
39 & 411.3178 & [17,13,7,2] & 3.379 & 411.3570 & [18,13,7,1] & 3.348\\
40 & 429.9292 & [17,13,8,2] & 3.428 & 430.0216 & [18,13,8,1] & 3.396\\
41 & 448.8795 & [17,14,8,2] & 3.470 & 449.0085 & [18,13,8,2] & 3.443\\
42 & 468.1613 & [17,14,8,3] & 3.517 & 468.2858 & [17,14,8,3] & 3.521\\
43 & 487.7576 & [17,14,9,3] & 3.560 & 487.8788 & [17,14,9,3] & 3.566\\
44 & 507.6915 & [18,14,9,3] & 3.573 & 507.7930 & [18,14,9,3] & 3.576\\
45 & 527.9573 & [18,15,9,3] & 3.613 & 528.0866 & [18,15,9,3] & 3.617\\
46 & 548.5782 & [19,15,9,3] & 3.624 & 548.6758 & [18,15,9,4] & 3.660\\
47 & 569.4976 & [18,15,10,4] & 3.697 & 569.5784 & [18,15,10,4] & 3.701\\
48 & 590.7302 & [19,15,10,4] & 3.708 & 590.8007 & [19,15,10,4] & 3.710\\
49 & 612.3010 & [19,16,10,4] & 3.746 & 612.3879 & [19,16,10,4] & 3.749\\
50 & 634.2032 & [20,16,10,4] & 3.756 & 634.2870 & [20,16,10,4] & 3.758\\
51 & 656.3994 & [19,16,11,5] & 3.822 & 656.4737 & [19,16,11,5] & 3.828\\
52 & 678.9044 & [19,15,11,6,1] & 3.858 & 678.9715 & [20,16,11,5] & 3.837\\
53 & 701.7045 & [18,17,11,6,1] & 3.920 & 701.8207 & [19,16,11,6,1] & 3.906\\
54 & 724.7968 & [18,17,12,6,1] & 3.957 & 724.9130 & [20,16,11,6,1] & 3.913\\
55 & 748.2062 & [19,17,12,6,1] & 3.969 & 748.3280 & [20,17,11,6,1] & 3.949\\
56 & 771.9241 & [20,17,12,6,1] & 3.981 & 772.0267 & [20,17,12,6,1] & 3.985\\
57 & 795.9622 & [20,18,12,6,1] & 4.014 & 796.0610 & [21,17,12,6,1] & 3.992\\
58 & 820.3013 & [21,18,12,6,1] & 4.023 & 820.4069 & [21,17,12,7,1] & 4.028\\
59 & 844.9824 & [22,18,12,6,1] & 4.031 & 845.0472 & [21,18,12,7,1] & 4.062\\
60 & 869.9167 & [21,18,13,7,1] & 4.094 & 869.9729 & [21,18,13,7,1] & 4.096\\
61 & 895.1684 & [22,18,13,7,1] & 4.101 & 895.2196 & [22,18,13,7,1] & 4.103\\
62 & 920.7048 & [21,17,14,8,2] & 4.155 & 920.7905 & [22,19,13,7,1] & 4.136\\
63 & 946.5161 & [20,19,14,8,2] & 4.212 & 946.6672 & [22,19,13,8,1] & 4.170\\
64 & 972.6175 & [21,19,14,8,2] & 4.222 & 972.7978 & [22,19,14,8,1] & 4.203\\
65 & 999.0138 & [20,19,14,9,3] & 4.270 & 999.2149 & [22,19,14,8,2] & 4.236\\
66 & 1025.691 & [20,20,14,9,3] & 4.305 & 1025.899 & [22,19,14,8,3] & 4.269\\
67 & 1052.661 & [20,20,15,9,3] & 4.338 & 1052.847 & [22,19,14,9,3] & 4.301\\
68 & 1079.908 & [21,20,15,9,3] & 4.345 & 1080.113 & [23,19,14,9,3] & 4.306\\
69 & 1107.457 & [22,20,15,9,3] & 4.354 & 1107.656 & [23,19,15,9,3] & 4.338\\
70 & 1135.298 & [23,20,15,9,3] & 4.363 & 1135.474 & [23,20,15,9,3] & 4.367\\
71 & 1163.410 & [21,21,15,10,4] & 4.429 & 1163.579 & [23,20,15,9,4] & 4.398\\
72 & 1191.798 & [21,21,16,10,4] & 4.465 & 1191.945 & [23,20,15,10,4] & 4.429\\
73 & 1220.463 & [22,21,16,10,4] & 4.472 & 1220.627 & [23,20,16,10,4] & 4.459\\
74 & 1249.421 & [23,21,16,10,4] & 4.481 & 1249.580 & [24,20,16,10,4] & 4.464\\
75 & 1278.653 & [23,21,14,11,5,1] & 4.503 & 1278.801 & [24,21,16,10,4] & 4.492\\
76 & 1308.157 & [22,21,16,11,5,1] & 4.554 & 1308.330 & [24,21,16,10,5] & 4.521\\
77 & 1337.925 & [22,22,16,11,5,1] & 4.583 & 1338.091 & [24,21,16,11,5] & 4.550\\
\hline
\end{tabular}
\end{table}
\vskip 5mm

\begin{table}
\centering
{\LARGE  Results for $78\le n \le 119$\par}
\begin{tabular} {@{\hspace{0.15cm}} c @{\hspace{0.5cm}} c @{\hspace{0.5cm}} r @{\hspace{0.5cm}} c @{\hspace{0.25cm}} | @{\hspace{0.25cm}} c @{\hspace{0.5cm}} r @{\hspace{0.5cm}} c @{\hspace{0.15cm}}}
\hline
$n$  & $E_{\rm MD}$ &  Configuration &  $R_{\rm avg}$  & $E_{\rm CM}$
 & Configuration & $R_{\rm ext}$\\
\hline
78 & 1367.963 & [22,22,17,11,5,1] & 4.612 & 1368.158 & [24,21,17,11,5] & 4.579\\
79 & 1398.252 & [22,22,17,11,6,1] & 4.640 & 1398.493 & [24,22,17,11,5] & 4.606\\
80 & 1428.827 & [22,22,17,12,6,1] & 4.669 & 1429.075 & [24,21,17,11,6,1] & 4.636\\
81 & 1459.672 & [22,22,18,12,6,1] & 4.696 & 1459.914 & [24,21,17,12,6,1] & 4.664\\
82 & 1490.795 & [23,22,18,12,6,1] & 4.702 & 1491.015 & [24,22,17,12,6,1] & 4.690\\
83 & 1522.175 & [23,23,18,12,6,1] & 4.727 & 1522.396 & [25,22,17,12,6,1] & 4.694\\
84 & 1553.845 & [24,23,18,12,6,1] & 4.733 & 1554.064 & [25,22,18,12,6,1] & 4.721\\
85 & 1585.785 & [24,24,18,12,6,1] & 4.757 & 1586.000 & [25,22,18,12,7,1] & 4.748\\
86 & 1618.019 & [25,24,18,12,6,1] & 4.764 & 1618.192 & [25,23,18,12,7,1] & 4.773\\
87 & 1650.497 & [24,24,18,13,7,1] & 4.812 & 1650.637 & [25,23,18,13,7,1] & 4.800\\
88 & 1683.217 & [24,24,19,13,7,1] & 4.837 & 1683.354 & [26,23,18,13,7,1] & 4.803\\
89 & 1716.199 & [24,24,19,13,7,2] & 4.860 & 1716.355 & [26,23,19,13,7,1] & 4.829\\
90 & 1749.437 & [24,24,32,6,2,2] & 4.888 & 1749.623 & [26,24,19,13,7,1] & 4.853\\
91 & 1782.929 & [24,24,19,14,8,2] & 4.912 & 1783.153 & [27,24,19,13,7,1] & 4.857\\
92 & 1816.674 & [24,24,19,14,8,3] & 4.934 & 1816.963 & [27,24,19,14,7,1] & 4.883\\
93 & 1850.697 & [24,24,19,14,9,3] & 4.957 & 1850.998 & [27,24,19,14,8,1] & 4.908\\
94 & 1884.964 & [24,24,20,14,9,3] & 4.984 & 1885.300 & [27,24,19,14,8,2] & 4.933\\
95 & 1919.486 & [25,25,19,14,9,3] & 4.990 & 1919.823 & [26,24,19,14,9,3] & 4.981\\
96 & 1954.254 & [25,25,20,14,9,3] & 5.013 & 1954.592 & [26,24,20,14,9,3] & 5.005\\
97 & 1989.285 & [25,25,20,15,9,3] & 5.038 & 1989.618 & [27,24,20,14,9,3] & 5.008\\
98 & 2024.569 & [25,25,21,15,9,3] & 5.061 & 2024.897 & [27,24,20,15,9,3] & 5.032\\
99 & 2060.131 & [26,25,21,15,9,3] & 5.066 & 2060.434 & [27,25,20,15,9,3] & 5.055\\
100 & 2095.931 & [26,26,21,15,9,3] & 5.088 & 2096.242 & [27,25,20,15,9,4] & 5.079\\
101 & 2131.979 & [26,26,21,14,10,4] & 5.112 & 2132.277 & [27,25,20,15,10,4] & 5.103\\
102 & 2168.273 & [26,26,20,16,10,4] & 5.136 & 2168.562 & [27,25,21,15,10,4] & 5.126\\
103 & 2204.812 & [26,26,21,16,10,4] & 5.158 & 2205.103 & [27,25,21,16,10,4] & 5.149\\
104 & 2241.601 & [26,26,22,16,10,4] & 5.180 & 2241.889 & [28,25,21,16,10,4] & 5.152\\
105 & 2278.670 & [27,26,22,16,10,4] & 5.185 & 2278.931 & [28,26,21,16,10,4] & 5.174\\
106 & 2315.960 & [26,26,21,16,11,5,1] & 5.222 & 2316.247 & [29,26,21,16,10,4] & 5.176\\
107 & 2353.486 & [26,26,22,15,11,6,1] & 5.249 & 2353.799 & [28,26,21,16,11,5] & 5.220\\
108 & 2391.253 & [26,26,20,17,12,6,1] & 5.273 & 2391.574 & [28,26,22,16,11,5] & 5.242\\
109 & 2429.264 & [26,26,21,17,12,6,1] & 5.294 & 2429.600 & [28,26,22,17,11,5] & 5.265\\
110 & 2467.525 & [26,26,21,18,12,6,1] & 5.315 & 2467.872 & [29,26,22,17,11,5] & 5.267\\
111 & 2506.023 & [27,27,20,18,12,6,1] & 5.320 & 2506.392 & [29,27,22,17,11,5] & 5.288\\
112 & 2544.754 & [27,27,21,18,12,6,1] & 5.341 & 2545.141 & [28,26,22,17,12,6,1] & 5.332\\
113 & 2583.739 & [27,27,22,18,12,6,1] & 5.361 & 2584.123 & [28,27,22,17,12,6,1] & 5.352\\
114 & 2622.965 & [27,27,23,18,12,6,1] & 5.382 & 2623.339 & [29,27,22,17,12,6,1] & 5.354\\
115 & 2662.432 & [27,27,24,18,12,6,1] & 5.402 & 2662.804 & [29,27,23,17,12,6,1] & 5.375\\
116 & 2702.166 & [28,28,23,18,12,6,1] & 5.406 & 2702.508 & [29,27,23,18,12,6,1] & 5.397\\
117 & 2742.111 & [28,28,24,18,12,6,1] & 5.426 & 2742.473 & [30,27,23,18,12,6,1] & 5.399\\
118 & 2782.346 & [29,28,24,18,12,6,1] & 5.430 & 2782.676 & [30,28,23,18,12,6,1] & 5.419\\
119 & 2822.794 & [29,29,24,18,12,6,1] & 5.449 & 2823.104 & [29,28,23,18,13,7,1] & 5.459\\
\hline
\end{tabular}
\end{table}
\vskip 5mm
\begin{table}
\centering
%\caption{\LARGE Results for $100\le N \le 149$}
{\LARGE Results for $120\le n \le 160$}
\begin{tabular} {@{\hspace{0.15cm}} c @{\hspace{0.5cm}} c @{\hspace{0.5cm}} r @{\hspace{0.5cm}} c @{\hspace{0.25cm}} | @{\hspace{0.25cm}} c @{\hspace{0.5cm}} r @{\hspace{0.5cm}} c @{\hspace{0.15cm}}}
\hline
$n$  & $E_{\rm MD}$ &  Configuration &  $R_{\rm avg}$  & $E_{\rm CM}$
 & Configuration & $R_{\rm ext}$\\
\hline
120 & 2863.472 & [28,28,24,19,12,8,1] & 5.489 & 2863.765 & [30,28,23,18,13,7,1] & 5.461\\
121 & 2904.384 & [28,28,41,14,6,2,2] & 5.511 & 2904.669 & [30,28,24,18,13,7,1] & 5.482\\
122 & 2945.520 & [28,28,25,17,14,8,2] & 5.531 & 2945.811 & [30,28,24,19,13,7,1] & 5.502\\
123 & 2986.892 & [28,28,25,32,6,2,2] & 5.550 & 2987.209 & [31,28,24,19,13,7,1] & 5.504\\
124 & 3028.501 & [28,28,42,14,6,3,3] & 5.571 & 3028.839 & [31,29,24,19,13,7,1] & 5.524\\
125 & 3070.333 & [29,29,25,16,14,9,3] & 5.574 & 3070.723 & [31,29,25,19,13,7,1] & 5.544\\
126 & 3112.398 & [29,29,22,20,14,9,3] & 5.594 & 3112.844 & [31,29,25,19,14,7,1] & 5.564\\
127 & 3154.706 & [29,29,25,18,14,9,3] & 5.613 & 3155.173 & [31,29,25,19,14,8,1] & 5.584\\
128 & 3197.232 & [29,29,23,20,15,9,3] & 5.633 & 3197.732 & [31,29,25,20,14,8,1] & 5.604\\
129 & 3240.000 & [29,29,24,20,15,9,3] & 5.652 & 3240.519 & [31,29,25,20,14,8,2] & 5.625\\
130 & 3282.996 & [29,29,24,21,15,9,3] & 5.671 & 3283.503 & [30,29,25,20,14,9,3] & 5.663\\
131 & 3326.229 & [29,29,25,21,15,9,3] & 5.690 & 3326.718 & [31,29,25,20,14,9,3] & 5.664\\
132 & 3369.693 & [29,29,26,21,15,9,3] & 5.708 & 3370.165 & [31,29,25,20,15,9,3] & 5.684\\
133 & 3413.392 & [30,30,26,20,15,9,3] & 5.711 & 3413.863 & [31,30,25,20,15,9,3] & 5.702\\
134 & 3457.312 & [30,30,26,21,15,9,3] & 5.730 & 3457.790 & [31,30,26,20,15,9,3] & 5.721\\
135 & 3501.457 & [30,30,24,21,16,10,4] & 5.750 & 3501.939 & [31,30,26,21,15,9,3] & 5.741\\
136 & 3545.825 & [30,30,27,19,16,10,4] & 5.768 & 3546.313 & [31,29,26,21,15,10,4] & 5.761\\
137 & 3590.431 & [30,30,27,20,16,10,4] & 5.786 & 3590.891 & [31,30,26,21,15,10,4] & 5.779\\
138 & 3635.262 & [30,30,26,22,16,10,4] & 5.805 & 3635.710 & [31,30,26,21,16,10,4] & 5.798\\
139 & 3680.315 & [30,30,27,22,16,10,4] & 5.822 & 3680.751 & [32,30,26,21,16,10,4] & 5.799\\
140 & 3725.602 & [30,30,28,22,16,10,4] & 5.840 & 3726.043 & [32,31,26,21,16,10,4] & 5.817\\
141 & 3771.105 & [30,30,2,44,17,11,6,1] & 5.861 & 3771.558 & [32,31,27,21,16,10,4] & 5.835\\
142 & 3816.828 & [31,31,23,22,17,11,6,1] & 5.864 & 3817.291 & [32,31,27,22,16,10,4] & 5.854\\
143 & 3862.771 & [30,30,26,22,17,11,6,1] & 5.896 & 3863.253 & [33,31,27,22,16,10,4] & 5.855\\
144 & 3908.934 & [31,31,25,22,17,11,6,1] & 5.899 & 3909.439 & [32,31,27,22,16,11,5] & 5.891\\
145 & 3955.315 & [30,30,26,23,17,12,6,1] & 5.932 & 3955.833 & [32,31,27,22,17,11,5] & 5.909\\
146 & 4001.923 & [31,31,25,23,17,12,6,1] & 5.936 & 4002.450 & [33,31,27,22,17,11,5] & 5.910\\
147 & 4048.753 & [31,31,25,23,18,12,6,1] & 5.953 & 4049.312 & [33,32,27,22,17,11,5] & 5.927\\
148 & 4095.803 & [30,30,28,23,18,12,6,1] & 5.984 & 4096.391 & [33,32,28,22,17,11,5] & 5.945\\
149 & 4143.077 & [31,31,26,24,18,12,6,1] & 5.989 & 4143.661 & [32,31,27,23,17,12,6,1] & 5.982\\
150 & 4190.571 & [30,30,29,24,18,12,6,1] & 6.019 & 4191.146 & [33,31,27,23,17,12,6,1] & 5.983\\
151 & 4238.292 & [31,31,28,24,18,12,6,1] & 6.022 & 4238.852 & [33,31,28,23,17,12,6,1] & 6.000\\
152 & 4286.238 & [32,32,27,24,18,12,6,1] & 6.025 & 4286.779 & [33,32,28,23,17,12,6,1] & 6.016\\
153 & 4334.404 & [32,32,28,24,18,12,6,1] & 6.041 & 4334.920 & [33,32,28,23,18,12,6,1] & 6.034\\
154 & 4382.789 & [32,32,29,24,18,12,6,1] & 6.058 & 4383.304 & [34,32,28,23,18,12,6,1] & 6.035\\
155 & 4431.402 & [32,32,30,24,18,12,6,1] & 6.074 & 4431.919 & [34,32,28,24,18,12,6,1] & 6.052\\
156 & 4480.240 & [33,33,29,24,18,12,6,1] & 6.076 & 4480.739 & [33,32,28,24,18,13,7,1] & 6.087\\
157 & 4529.280 & [33,33,30,24,18,12,6,1] & 6.092 & 4529.763 & [34,32,28,24,18,13,7,1] & 6.087\\
158 & 4578.533 & [31,31,53,19,14,6,2,2] & 6.140 & 4579.002 & [34,32,29,24,18,13,7,1] & 6.104\\
159 & 4627.983 & [32,32,30,41,14,6,2,2] & 6.144 & 4628.461 & [34,33,29,24,18,13,7,1] & 6.120\\
160 & 4677.653 & [32,32,53,19,14,6,2,2] & 6.159 & 4678.133 & [34,33,29,24,19,13,7,1] & 6.137\\
\hline
\end{tabular}
\end{table}

\begin{table}
\centering
%\caption{\LARGE Results for $166\le N \le 200$}
{\LARGE Results for $161\le n \le 200$}
\begin{tabular} {@{\hspace{0.15cm}} c @{\hspace{0.5cm}} c @{\hspace{0.5cm}} r @{\hspace{0.5cm}} c @{\hspace{0.25cm}} | @{\hspace{0.25cm}} c @{\hspace{0.5cm}} r @{\hspace{0.5cm}} c @{\hspace{0.15cm}}}
\hline
$n$  & $E_{\rm MD}$ &  Configuration &  $R_{\rm avg}$  & $E_{\rm CM}$
 & Configuration & $R_{\rm ext}$\\
\hline
161 & 4727.551 & [33,33,30,41,14,6,2,2] & 6.163 & 4728.043 & [35,33,29,24,19,13,7,1] & 6.138\\
162 & 4777.653 & [32,32,54,20,14,6,2,2] & 6.192 & 4778.179 & [35,33,29,25,19,13,7,1] & 6.155\\
163 & 4827.973 & [32,32,55,20,14,6,2,2] & 6.208 & 4828.523 & [35,33,30,25,19,13,7,1] & 6.171\\
164 & 4878.503 & [31,31,31,25,20,14,9,3] & 6.237 & 4879.078 & [35,34,30,25,19,13,7,1] & 6.187\\
165 & 4929.243 & [32,32,55,20,14,6,3,3] & 6.241 & 4929.869 & [36,34,30,25,19,13,7,1] & 6.188\\
166 & 4980.198 & [31,31,31,25,21,15,9,3] & 6.270 & 4980.862 & [35,34,30,25,19,14,8,1] & 6.221\\
167 & 5031.363 & [32,32,31,45,15,6,3,3] & 6.274 & 5032.046 & [35,34,30,25,20,14,8,1] & 6.237\\
168 & 5082.740 & [32,32,31,26,20,15,9,3] & 6.288 & 5083.459 & [35,34,30,25,20,14,8,2] & 6.254\\
169 & 5134.328 & [32,32,31,26,21,15,9,3] & 6.305 & 5135.059 & [34,34,30,25,20,14,9,3] & 6.287\\
170 & 5186.133 & [32,32,32,26,21,15,9,3] & 6.320 & 5186.850 & [35,34,30,25,20,14,9,3] & 6.287\\
171 & 5238.158 & [33,33,57,21,15,6,3,3] & 6.323 & 5238.859 & [35,34,30,25,20,15,9,3] & 6.303\\
172 & 5290.383 & [33,33,31,27,21,15,9,3] & 6.338 & 5291.081 & [35,34,30,26,20,15,9,3] & 6.319\\
173 & 5342.830 & [33,33,32,27,21,15,9,3] & 6.354 & 5343.522 & [35,34,31,26,20,15,9,3] & 6.335\\
174 & 5395.482 & [32,32,32,26,22,16,10,4] & 6.382 & 5396.169 & [35,34,31,26,21,15,9,3] & 6.351\\
175 & 5448.340 & [33,64,26,22,16,10,2,2] & 6.385 & 5449.031 & [36,34,31,26,21,15,9,3] & 6.352\\
176 & 5501.406 & [33,33,32,27,21,16,10,4] & 6.402 & 5502.099 & [36,35,31,26,21,15,9,3] & 6.366\\
177 & 5554.678 & [33,33,32,27,22,16,10,4] & 6.416 & 5555.369 & [35,35,31,26,21,15,10,4] & 6.398\\
178 & 5608.165 & [33,33,33,27,22,16,10,4] & 6.431 & 5608.845 & [36,35,31,26,21,15,10,4] & 6.398\\
179 & 5661.871 & [34,34,59,22,16,10,2,2] & 6.434 & 5662.526 & [36,35,31,26,21,16,10,4] & 6.414\\
180 & 5715.769 & [34,34,32,28,22,16,10,4] & 6.449 & 5716.419 & [36,35,31,27,21,16,10,4] & 6.430\\
181 & 5769.872 & [33,33,2,31,47,17,11,6,1] & 6.480 & 5770.531 & [36,35,32,27,21,16,10,4] & 6.445\\
182 & 5824.177 & [33,33,33,28,20,17,11,6,1] & 6.494 & 5824.842 & [36,35,32,27,22,16,10,4] & 6.461\\
183 & 5878.689 & [33,33,60,5,17,17,11,6,1] & 6.506 & 5879.372 & [36,36,32,27,22,16,10,4] & 6.475\\
184 & 5933.401 & [34,34,34,28,18,17,12,6,1] & 6.514 & 5934.105 & [37,36,32,27,22,16,10,4] & 6.475\\
185 & 5988.323 & [34,34,1,57,23,17,12,6,1] & 6.527 & 5989.067 & [36,36,32,27,22,16,11,5] & 6.506\\
186 & 6043.450 & [34,34,34,25,23,17,12,6,1] & 6.542 & 6044.200 & [36,36,32,27,22,17,11,5] & 6.522\\
187 & 6098.766 & [34,34,34,24,24,18,12,6,1] & 6.559 & 6099.538 & [37,36,32,27,22,17,11,5] & 6.522\\
188 & 6154.300 & [34,34,34,25,24,18,12,6,1] & 6.572 & 6155.080 & [37,36,32,28,22,17,11,5] & 6.537\\
189 & 6210.038 & [34,34,34,26,24,18,12,6,1] & 6.586 & 6210.839 & [37,36,32,28,23,17,11,5] & 6.552\\
190 & 6265.982 & [34,34,34,27,24,18,12,6,1] & 6.600 & 6266.794 & [37,36,33,28,23,17,11,5] & 6.567\\
191 & 6322.128 & [34,34,34,28,24,18,12,6,1] & 6.614 & 6322.955 & [36,36,32,28,23,17,12,6,1] & 6.598\\
192 & 6378.489 & [34,34,34,29,24,18,12,6,1] & 6.628 & 6379.295 & [37,36,32,28,23,17,12,6,1] & 6.597\\
193 & 6435.051 & [35,35,34,28,24,18,12,6,1] & 6.630 & 6435.837 & [37,36,33,28,23,17,12,6,1] & 6.612\\
194 & 6491.807 & [35,35,34,29,24,18,12,6,1] & 6.644 & 6492.585 & [37,36,33,28,23,18,12,6,1] & 6.627\\
195 & 6548.773 & [35,35,35,29,24,18,12,6,1] & 6.658 & 6549.548 & [37,37,33,28,23,18,12,6,1] & 6.641\\
196 & 6605.955 & [35,35,35,30,24,18,12,6,1] & 6.672 & 6606.711 & [37,37,33,29,23,18,12,6,1] & 6.655\\
197 & 6663.328 & [36,36,34,30,24,18,12,6,1] & 6.674 & 6664.081 & [38,37,33,29,23,18,12,6,1] & 6.655\\
198 & 6720.909 & [36,36,35,30,24,18,12,6,1] & 6.688 & 6721.650 & [38,37,33,29,24,18,12,6,1] & 6.670\\
199 & 6778.722 & [35,1,35,35,30,44,12,6,1] & 6.710 & 6779.422 & [38,37,34,29,24,18,12,6,1] & 6.684\\
200 & 6836.692 & [35,35,35,53,19,13,6,2,2] & 6.729 & 6837.393 & [37,37,34,29,24,18,13,7,1] & 6.714\\
\hline
\end{tabular}
\end{table}
\end{widetext}
\end{document}